\documentclass[10pt]{IEEEtran}

\usepackage{booktabs}

\usepackage{enumitem}
\usepackage{url}
\usepackage[pdftex]{graphicx}
\usepackage{float}
\usepackage{epstopdf}
\usepackage{amsmath}
\usepackage{amssymb}
\usepackage{pifont}
\usepackage[table,xcdraw]{xcolor}
\usepackage{array}
\usepackage{longtable}
\usepackage{supertabular}
\usepackage{cite}
\usepackage{moresize}
\usepackage{caption}
\usepackage{subcaption}
\usepackage{color}

\begin{document}

\title{Exploiting the power of multiplicity: \\a holistic survey of network-layer multipath}
\author{Junaid Qadir, Anwaar Ali, Kok-Lim Alvin Yau, Arjuna Sathiaseelan, Jon Crowcroft 
\thanks{Junaid Qadir is an Assistant Professor at the Electrical Engineering Department of SEECS, NUST. Anwaar Ali is a postgraduate student of Electrical Engineering at the Electrical Engineering Department of SEECS, NUST. Kok-Lim Alvin Yau is an Associate Professor at the Faculty of Science and Technology, Sunway University, Malaysia. Arjuna Sathiaseelan is a Senior Research Associate at the Computer Laboratory, University of Cambridge. Jon Crowcroft is the Marconi Professor of Networked Systems at the Computer Laboratory of the University of Cambridge.}
}
\maketitle

\begin{abstract}

The Internet is inherently a multipath network---for an underlying network with only a single path connecting various nodes would have been debilitatingly fragile. Unfortunately, traditional Internet technologies have been designed around the restrictive assumption of a single working path between a source and a destination. The lack of native multipath support constrains network performance even as the underlying network is richly connected and has redundant multiple paths. Computer networks can exploit the power of multiplicity---through which a diverse collection of paths is resource pooled as a single resource---to unlock the inherent redundancy of the Internet. This opens up a new vista of opportunities promising increased throughput (through concurrent usage of multiple paths) and increased reliability and fault-tolerance (through the use of multiple paths in backup/ redundant arrangements). There are many emerging trends in networking that signify that the Internet's future will be unmistakably multipath,  including the use of multipath technology in datacenter computing; multi-interface, multi-channel, and multi-antenna trends in wireless; ubiquity of mobile devices that are multi-homed with heterogeneous access networks; and the development and standardization of multipath transport protocols such as MP-TCP. 

The aim of this paper is to provide a comprehensive survey of the literature on network-layer multipath solutions. We will present a detailed investigation of two important design issues, namely the \textit{control plane problem} of how to compute and select the routes, and the \textit{data plane problem} of how to split the flow on the computed paths. The main contribution of this paper is a systematic articulation of the main design issues in network-layer multipath routing along with a broad-ranging survey of the vast literature on network-layer multipathing. We also highlight open issues and identify directions for future work.


\end{abstract}

\begin{IEEEkeywords}
Multipath, Diversity, Routing, Load Balancing.
\end{IEEEkeywords}

\section{Introduction}
\label{sec:intro}

It is common for communication networks to be built with significant redundancy for the purpose of fault tolerance and reliability. This results in the availability of multiple paths that can be used for connecting network nodes. These multiple paths can potentially be resource pooled for the purpose of routing and forwarding packets, and can be used either simultaneously or in a backup configuration, for increased reliability, end-to-end throughput, network efficiency, and fault tolerance. Unfortunately, despite the widespread path diversity available on the Internet, traditional Internet technologies have focused mainly on single-path routing for the sake of its simplicity and lower overhead. This has led to the artificial lockup of Internet's capacity, lower fault tolerance and reliability, and inflexible support for quality of service (QoS). This is even in contrast to telephone networks, which have traditionally adopted multiple-path routing\footnote{Traditional circuit-switched public switched telephone networks (PSTN) use a type of multipath routing known as \textit{hierarchical routing} or \textit{alternative path routing}. Each exchange typically maintains precomputed routing tables with multiple alternative routes to any given destination with a fixed first-choice route, and one or more alternative routes hierarchically arranged for if the first-choice route is blocked, from which the exchange selects the route at the time of the call depending on network congestion level \cite{bertsekas1992data}.} since it leads to better reliability and greater customer satisfaction (due to the lesser probability of call blocking) \cite{keshav1997engineering}.


Notwithstanding the lack of native multipath support on the current Internet, there is a growing convergence in the Internet community that multipath routing and forwarding, or \textit{multipathing} in short, will play a big role in the future Internet by removing the artificial constraints of single-path routing. 
As we shall discuss next, single-path based approaches are deficient on many accounts.


%

\subsection{Deficiencies of Shortest Single-Path Routing}
\label{subsec:deficiencies}

There are two fundamental drawbacks that traditional routing techniques based on shortest single-path routing suffer.

\vspace{2mm}
\textit{Firstly}, it potentially limits the network throughput since only one path is accessible per source-destination pair. It has been shown that single-path routing is constraining optimal routing solutions, and that elegant optimal routing solutions can be developed if multiple paths with flexible splitting are utilized \cite{wang2011cost}. The problem of joint optimization of routing and congestion control which belongs to the category of the general multi-commodity flow (MCF) problem is considered in \cite{wang2009howbad} for both single-path and multi-path configurations. It is shown that this problem is \textit{non-convex} and NP-hard when the user can utilize a \textit{single path} only. On the other hand, if all available paths can be used with \textit{source-based multipath} routing, the problem is convex and admits an elegant solution. Wang et al. \cite{wang2009howbad} go on to investigate the performance loss of being restricted to a single path. This performance penalty of not using multiple paths is referred to as `the cost of not splitting' \cite{wang2011cost, wang2009howbad}. This performance penalty, although hard to determine (being an NP-hard problem), is known to be strictly positive. 

\vspace{2mm}
\textit{Secondly}, it is susceptible to routing oscillations since any changes in link lengths can cause abrupt traffic shifts. It is well-known in the literature that adaptive, shortest-path routing protocols---especially those that rely on link delays---can result in oscillatory behavior and instability leading to performance degradation \cite{wang1992analysis, bertsekas1987data}.  An example of an early delay-based adaptive routing algorithm is the distance-vector ARPANET routing protocol \cite{mcquillan1977arpa} that was later found out to be unstable \cite{mcquillan1980new, khanna1989revised}. 


\vspace{2mm}
Optimal routing based on the rigorous mathematical theory of the MCF problem was proposed to eliminate several disadvantages of single-path routing by optimizing an average delay-like performance measure \cite{bertsekas1992data}. \textit{Optimal routing is essentially a multipath routing framework}. In optimal routing, traffic from a source-destination pair is split at strategic points allowing gradual shifting of traffic among alternative paths. Optimal routing aims to send traffic exclusively on paths that are shortest with respect to some link lengths computed on the basis of flows carried on those links. In optimal routing, for low input traffic, a source-destination pair tends to use only one path (which is the fastest in terms of packet transmission time), while for increasing input traffic, additional paths are utilized to avoid overloading the shortest path.

\subsection{Why the Internet's Future is Unmistakably Multipath?}
\label{sec:whyFuture}

There are five main trends that indicate that the future Internet is going to be intimately intertwined with multipathing.

\vspace{1mm}
\subsubsection{Datacenters Trend} With the increasingly central role of \textit{datacenters (DCs)} in the modern Internet, multipathing is becoming an explicitly designed feature of modern architecture. Unlike the traditional networking topologies that have evolved spontaneously, DCs are highly engineered architectural artifacts that have intrinsically incorporated multipathing for increased performance and reliability. 

\vspace{1mm}
\subsubsection{Wireless Trend} In wireless communication, multipath plays an important role at various layers of the protocol stack (including the transport, network, and physical layers). Apart from the importance of multipathing at the transport and network layers due to the multihoming of modern mobile devices (such as smartphones), multipath also plays a critical role at the physical layer. Once considered as an antagonist of efficient wireless communications, due to the deleterious effects of multipath fading in which the reflected echoes of the transmission cancel each other destructively affecting successful reception, multipath is now considered advantageous at the physical layer with the recent advances in adaptive signal processing and antenna techniques in multiple input and multiple output (MIMO) technologies. To keep up with the growing bandwidth requirements of modern applications, almost all of the emerging wireless technologies rely on the modern trend of equipping wireless nodes with MIMO antennas. With multiple antennas relying on the extra spatial degrees of freedom, the ability to successfully use multiple paths has become critical to the success of modern wireless networks.

\vspace{1mm}
\subsubsection{Multiple Wireless NICs Trend} Modern wireless network technologies have also made it feasible to equip network nodes with multiple network interface cards (NICs) tuned to orthogonal channels. The ready availability of commodity NICs, and the availability of modern networks to support multiple interfaces, has opened up the possibility of the successful exploitation of multipath technology to improve network performance.

\vspace{1mm}
\subsubsection{Multihoming with Heterogeneous Technologies} 

Multihoming is a technique that has been traditionally used by server farms on the Internet for reliability and high performance. Analysis of multihoming has shown significant performance benefits that accrue with multhihoming, with a careful selection of three or more Internet service providers (ISPs) providing performance improvement, in terms of transfer throughput and round-trip-time (RTT), of upto 25\% \cite{akella2003measurement}. While traditionally multihoming has been relevant only for critical servers, it has now become a mainstream with billions of mobile devices now multihomed with heterogeneous access technologies (such as Wi-Fi/ 3G). A wireless device (such as a smartphone) can exploit its interfaces to heterogeneous technologies in order to improve its reliability, throughput, and fault tolerance. In recent times, there have been multipath transport-layer initiatives that are built specifically to incorporate such multihoming. In particular, the standardization of MP-TCP \cite{Paasch:2014:MT:2580723.2578901} in RFC6824 has led to mainstream applications of multipath: e.g., the Siri intelligent personal assistant on Apple's iOS7 uses MP-TCP to increase efficiency and reliability.  



\vspace{1mm}
\subsubsection{Path Diversity in ISPs}  Teixeira et al. have studied the availability of \textit{path diversity} (the availability of multiple possible paths between two communicating entities) in ISPs \cite{teixeira2003search}. It was shown that a typical large ISP\footnote{Empirical calculations were performed on a tier-1 Sprint ISP network using heuristics for inferring path diversity based on active measurements using the \textit{Rocketfuel} \cite{spring2002measuring} toolset.} has a significant path diversity. The abundant availability of path diversity does not necessarily imply that the best paths are adopted though---in fact, statistics show that at any time 30 to 80 percent of the Internet traffic chooses a single suboptimal path even though alternative low-loss lesser-delay paths exist \cite{savage1999end}. With the inherent path diversity in ISPs, there is a strong interest in the service provider community to exploit multiple paths \cite{he2008toward}.

\subsection{Benefits of Multipathing}
\label{subsec:benefits}

We provide a description of some of the major applications and features of multipathing below:

\vspace{1mm}
\subsubsection{Resource Pooling}  Resource pooling is a broad Internet architectural design principle---focusing on improving resource efficiency by presenting a collection of resources as a single pooled resource---that has many manifestations. For example, some of the most disruptive innovations on the Internet---packet switching, content-delivery networks (CDN), peer-to-peer networks (P2P), cloud computing---are based on resource pooling \cite{wischik2008resource}. The packet switching technology, which acts as a foundation of the Internet, is essentially a resource pooling technique. To put things in perspective, circuit switching techniques use fixed multiplexing schemes in which dedicated and isolated non-pooled circuits are utilized. In contrast, statistical multiplexing---the underlying technique of packet switching---achieves resource pooling by allowing a burst of traffic on a single circuit to use spare capacity on other circuits. CDNs and P2P networks utilize resource pooling to ensure efficient network operations. P2P protocols such as BitTorrent use swarming downloads that let receivers pool together multiple peers as a data source thereby using multipath to pool the network paths to these peers. Cloud computing relies on pooled resources gathered at central DCs to support the computing requirements of multiple tenants through virtualization. Multipathing has been envisioned as `\textit{Packet Switching 2.0}' \cite{wischik2008resource} since it allows an analogous resource pooling benefit at the network layer. In particular, multipath routing can pool together a number of distinct links to provide the abstraction of a single unified network resource. With the abstraction of a single resource pool, the chances of a demand being refused due to one's resources being utilized reduces. 

\vspace{1mm}
\subsubsection{Load Balancing} 

With resource pooling, a collection of resources acts as a single virtual resource. The biggest promise of resource pooling is load balancing that allows congestion control techniques for diffusing congestion over resources by equitably distributing the load amongst the elements of a resource pool. This helps to avoid situations where a certain network resource unnecessarily acts as a bottleneck although the network has alternative resources that can help to relieve the congestion \cite{Valiant1990, zhang2010valiant}. Load balancing can be done over resources (e.g., the load balancing over multiple access links and paths as done in multihoming and multipath transport respectively) or can be temporal (i.e., the load is balanced between peak and non-peak hours to exploit the typical diurnal network traffic pattern). Load balancing over time shifts the transmission of data from peak time to off-peak time and can help in reducing costs such as the commonly used 95 percentile pricing scheme \cite{clegg2014tardis}.

\vspace{1mm}
\subsubsection{Efficiency} 

Another major benefit of resoure pooling through multiple paths is efficiency. It is well known in the literature that the Internet traffic is bursty with a pronounced correlation with a large variance of the traffic volume over long time periods \cite{paxson1995wide}, \cite{leland1994self}. Allocating fixed amount of resources to individual circuits is inefficient due to the bursty nature of the Internet traffic, which can result in largely idle resources or resources that are insufficiently provisioned. One important benefit associated with resource pooling is that it leads to increased efficiency. Resource pooling, or statistical multiplexing of shared resources, suits the Internet traffic well. This is because the law of averaging (or the law of large numbers) does not apply to individual flows. Resource pooling is useful since it exploits the statistical regularity of the aggregate of individual bursty circuits. This results in efficient utilization of resources without extensive over-provisioning.

\vspace{1mm}
\subsubsection{Reliability and Fault Tolerance} 

With the migration of critical businesses to the Internet, the reliability and availability of the Internet has become extremely important. Fault tolerance mechanisms for networks without multipath support are unwieldy with a link/ node failure potentially resulting in significant recovery times. The resource pooling benefit of multipathing allows the underlying network to gracefully handle loss of capacity or the failure of individual links/ paths through diffusing this anomaly by shifting traffic to the working members of the resource pool. This allows enhanced fault tolerance and helps to build more resilient networks.

Network communication can utilize multipathing to enhance its \textit{reliability} at various layers \cite{bailis2014network}. At the network layer, we can have multihomed servers that have different ISP connections; another option is to utilize backup paths with traffic being switched to an alternative path in the case of failure of the primary path. At the physical layer, techniques such as antenna diversity can be used to combat fading effects in wireless communication. 

Multipathing can also be used to facilitate network security. More specifically, the use of multiple radio interfaces has been proposed for achieving greater data \textit{confidentiality} by encrypting data and splitting it into multiple parts that are then transmitted using different physical channels \cite{ye2008improving}. Multipathing can also be used to mitigate malicious adversaries that drop packets intentionally on a path by circumventing the adversary using an alternative disjoint path if available. 


\vspace{1mm}
\subsubsection{Higher Throughput} A great number of modern high-performance Internet technologies utilize multipathing in one way or another. At the physical layer, MIMO technology is revolutionizing the wireless industry by promising significantly enhanced communication rates. The technique of space-division multiple access (SDMA), used by multi-user MIMO (MU-MIMO), allows multiple transmitters to send---and multiple receivers to receive---separate signals simultaneously over the same band. The improvement offered through multipathing at the link layer and the network layer is dramatic in environments where the traffic is concentrated on only a few links while other perfectly fine links are not used at all (e.g., in bridged networks where the spanning tree protocol is used). Other high-performance proposals include multipath-TCP (MP-TCP) at the transport layer that allows the pooling of multiple links into a pooled connection to support higher speed: for example, there has been a recent demonstration of a transfer achieving a goodput of 50 Gbps using MP-TCP with nodes equipped with 6-10 Gbps commodity NICs \cite{fastestTCP}.

\vspace{1mm}
\subsubsection{Flexibility due to Pluralism and Multiplicity}

In recent literature \cite{kahneman2011thinking}, the metaphors of ``the hedgehog and the fox''\footnote{Based on a reported saying from the ancient Greek poet Archilochus who had said: ``\textit{the fox knows many things but the hedgehog knows one big thing}''.}---popularized by Isiah Berlin in an essay of the same name---are being used to highlight the importance of flexibility that accrues with pluralism. For example, it has been shown that a forecaster who systematically entertains a variety of possibilities outperforms a rigid one-dimensional forecaster fixated on a single potential interpretation \cite{tetlock2005expert}. There is also a wealth of literature that supports the claim of ``\textit{the wisdom of crowds}'' \cite{surowiecki2005wisdom}. 

Many of the benefits of multipathing are brought about by the \textit{`diversities'} it offers. The benefits of diversity have been established in a broad spectrum of fields ranging from communications, biology, economics, to sociology. Intuitively speaking, diversity is useful in a variety of fields since it allows hedging in uncertain situations to minimize risk and thereby downplays the negative effects of uncertainty.


\vspace{1mm}
\subsubsection{Support for Multimedia Applications} The modern rise of high bandwidth multimedia applications can benefit significantly from the resilience offered by multipath techniques. While the traditional packet-switched Internet is inherently a best-effort network where a single path is mostly in use, multiple paths can coexist at the same time. As scarce bandwidth and packet loss are more significant problems for today's streaming media because of its increased transmission requirements, concurrent transmission of multimedia applications over multiple paths can be adopted to resolve the aforementioned problems.

\vspace{1mm}
\subsubsection{Role in the Future Internet Architectures} 

It is anticipated that multipathing will play a major role in the future Internet architectures and next-generation wireless and mobile networks. It is already typical for mobile phones to be multihomed with heterogeneous connectivity (e.g., Wi-Fi and 3G). The upcoming 5G standard (expected to come to fruition around 2020) is expected to integrate and resource pool a number of diverse access technologies \cite{andrews2014will} with multipathing playing its part in seamless handovers and interworking of heterogeneous technologies. With the resource pooling of cellular and Wi-Fi networks, the possibilities of Wi-Fi offloading and cellular onloading emerge. With Wi-Fi offloading, the performance and scalability of mobile networks can be improved by offloading data traffic to the Wi-Fi networks \cite{gonzalez2013radio}. With cellular onloading, on the other hand,  home broadband/ Wi-Fi connections are augmented with cellular data services for bandwidth aggregation \cite{vallina2012david}. Apart from wireless networks, multipathing is also useful in emerging networking configurations (such as DCs and clouds) and in other upcoming Internet architectures such as P2P, CDN, software-defined network (SDN), and content-centric network (CCN) \cite{carofiglio2013multipath}, \cite{rossini2013evaluating}, \cite{tyson2012trace}. 

\begin{figure*}
\centering
\includegraphics[width=.75\textwidth]{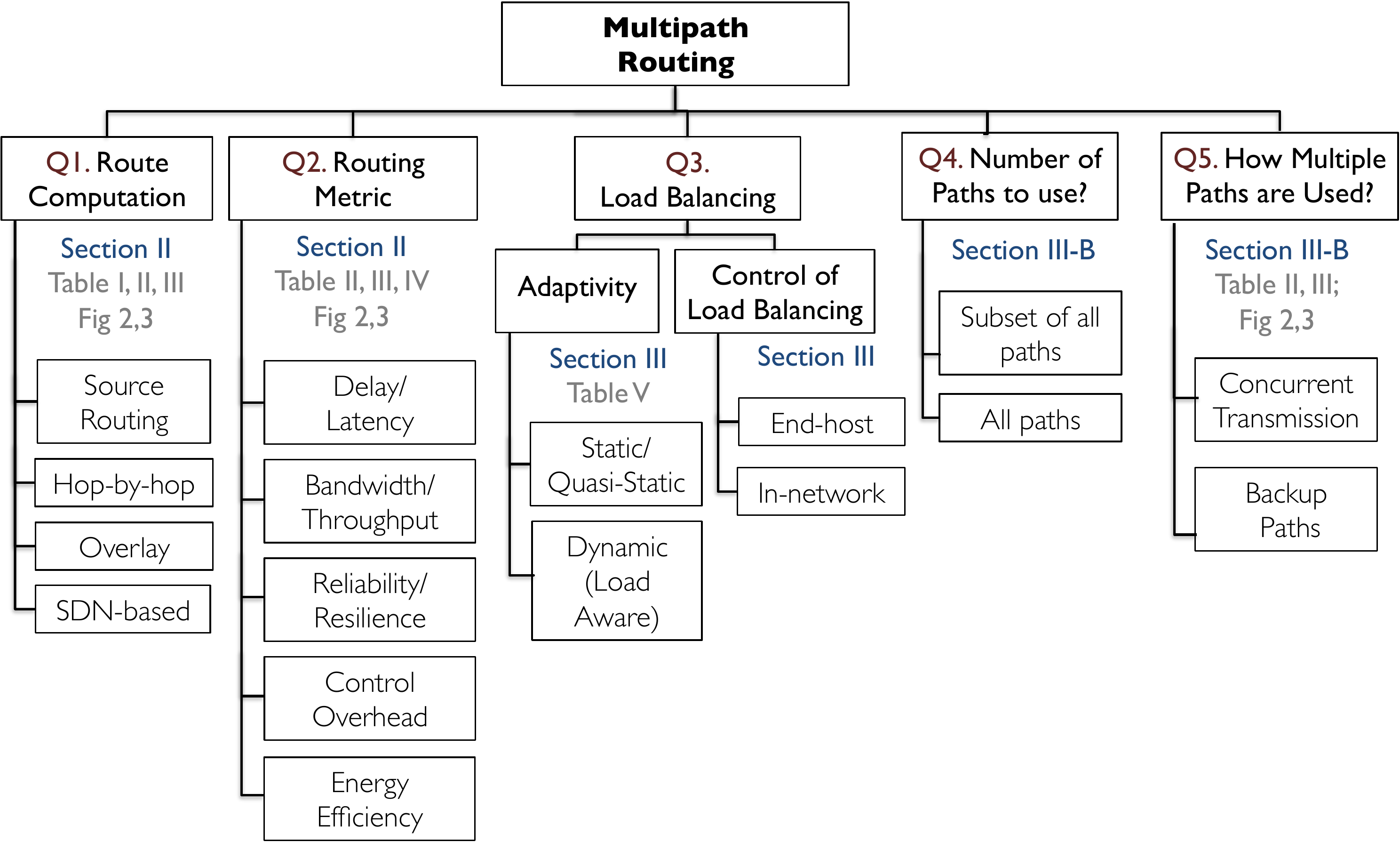}
\caption{Various network-layer multipathing design choices (and their coverage in this paper).}
\label{fig:comrpro}
\end{figure*}

\vspace{1mm}
\subsubsection{Context-aware Services} It should be noted that different networking applications have different needs. In particular, various application configurations can coexist: e.g., the first application prefers a low-delay path, the second prefers a high-throughput connection, and another prefers a reliable connection. Using multipathing, context-aware services are in line with the application requirements.

\subsection{Challenges in Implementing Multipath Routing}
\label{subsec:ciimrou}

Even though the future of the Internet promises to be intertwined with multipath routing, there are numerous challenges that need to be surmounted. A major challenge in implementing Internet-wide multipath routing is to achieve scalability due to the resulting considerable computational and storage overhead. Specifically, multipath routing entails significant overhead both in the control as well as the data plane. Control plane overhead refers to the need of additional bandwidth and processing resources for exchanging path information, while the data plane overhead refers to the large forwarding table and the higher memory requirement associated with forwarding entries corresponding to an increased number of paths. Another stumbling block is the lack of complete control of any organizations over the entire end-to-end path. We will discuss more open and current research issues later in the paper in section \ref{sec:future_wrk}.

\subsection{Organization and Contributions of This Paper}

We organize this paper around five fundamental design issues that relate to network-layer multipathing. These designs issues are presented below in the form of design questions:

\begin{enumerate}[label=\textbf{Q\arabic*}.,leftmargin=*]

\item \textit{Route computation}: How to compute the multiple paths to be used?

\item \textit{Route metric:} What routing metrics should be used to construct the `best' multiple paths?

\item \textit{Load balancing/ Flow splitting:} How to have a traffic flow use the constructed paths?

\item \textit{Number of paths:} How \textit{many} paths should be constructed and used?

\item \textit{Concurrent usage:} Should the multiple paths be used concurrently or as backups?

\end{enumerate}

The overall organization of this paper is illustrated in Figure \ref{fig:comrpro}. The main questions that we analyze, and present solutions for, have already been presented. In Section \ref{sec:cons_sel_of_routes}, we discuss various techniques that can be used for the construction and selection of multiple paths (Q1) and also the metrics that are used for facilitating route construction (Q2). In Section \ref{sec:usage_mp}, we tackle the issue of how to use the multiple constructed paths (Q3). In particular, we discuss issues relating to load balancing, flow splitting, and multipath congestion control. We discuss the choice of how many paths to use (Q4) in Section \ref{sec:con_backup}, and whether to adopt backup or concurrent multipathing (Q5) in Section \ref{sec:con_backup}. In Section \ref{sec:future_wrk}, we present potential open research issues in the field of multipathing at the network layer followed by our main conclusions in Section \ref{sec:con}.

The main contribution of this work is a comprehensive survey of network-layer multipath routing in which apart from our network-layer focus we also provide a balanced and holistic cross-layer coverage of multipathing (recognizing the fact that many important issues in multipathing---such as resource pooling, congestion control, flow splitting---involve cross-layer interaction. While there exist quite a few survey papers on multipathing protocols, these papers are typically confined to particular networking configurations (such as wireless sensor networks  \cite{radi2012multipath}, \cite{sha2013multipath}, \cite{tarique2009survey}, mobile ad hoc networks \cite{mueller2004multipath}, \cite{lou2006performance}, \cite{parissidis2006multi}, etc.) We will present a more general paper that will comprehensively describe all the important multipath proposals/ ideas and will also tie them up with new advances in transport-layer multipathing. With the increasing importance of multipathing on the Internet, there is a strong need to have such a paper to benefit practitioners and researchers in the networking community. This paper is timely due to the huge interest in this domain (emerging from new research/ standardization work on multipath protocols, such as MP-TCP). The success of higher-layer multipath protocols depends critically on support from lower-layer protocols (in particular, the network-layer protocol). 
 
\vspace{2mm}
\section{How to Construct Routes?}
\label{sec:cons_sel_of_routes}

On the Internet, the task of the construction of routes is handled by the \textit{control plane}, while the task of forwarding the packets based on the computed routes is performed by the Internet's \textit{data plane}. The Internet's control plane is typically managed through dynamic routing protocols. Generally speaking, it is desirable for a routing protocol to (i) choose `optimal' paths, (ii) be robust to dynamic network topology in order to avoid instability in the form of oscillatory route flaps, and (iii) minimize routing overheads in terms of memory required, messages exchanged, etc. There are several choices in the design of multipath routing protocols---including centralized or distributed routing, static or dynamic routing, source-based or hop-by-hop routing---corresponding to different tradeoffs with respect to complexity and resource usage. 

In this section, we focus on the various techniques proposed in the literature for the computation of routes. These techniques include ``source routing'' in which the end-to-end route is computed and specified by a source and ``hop-by-hop routing'' in which network routers exchange control packets to establish state that is then used for hop-by-hop forwarding. Broadly speaking, the various multipath routing techniques can be classified into four categories of: (i) source routing, (ii) hop-by-hop routing, (iii) overlay routing, and (iv) SDN-based routing. In the following subsections, we discuss each of these categories in this order. A summary of the multipath proposals presented in this section---along with a description of their route computation technique, routing metrics adopted, and specific networking configuration assumed---is presented in Tables \ref{tab:Determining_Routes}, \ref{tab:MultipathRouting}, and \ref{tab:MultipathRoutingWireless}.

\begin{table*}[!ht]
\centering
\ssmall
\caption{Frameworks for Determining Multiple Paths (Route Selection)}
\label{tab:Determining_Routes}
\begin{tabular}{p{5.5cm}p{1cm}p{10cm}}
\hline

\cellcolor[HTML]{EFEFEF}\textbf{\textit{Project}}
& \cellcolor[HTML]{EFEFEF}\textbf{\textit{Year}} &
\cellcolor[HTML]{EFEFEF}\textbf{\textit{Description}}  \\
\hline

\\
\multicolumn{2}{l}{\textbf{\underline{\emph{Source Routing}}}}\\
\\

Split Multipath Routing (SMR) \cite{nasipuri1999demand}, \cite{wang2000multipath} & 1999, 2000 & Applies a probing-based mechanism to achieve a balanced load over multiple constructed paths. \\

Multipath MPLS \cite{seok2001dynamic} & 2001 & Constructs multiple paths in polynomial time subject to a constraint (i.e., number of hops and paths). \\

Feedback based routing (FBR)  \cite{zhu2003feedback} & 2003 & Performs routing decision at the access router with two routes computed for each network prefix using dynamic quality monitoring based on structural information disseminated by the transit routers .\\

BANANAS \cite{kaur2003bananas} & 2003 & Enables multipath routing in conventional routing protocols (e.g., OSFP and BGP) using global identifiers for abstraction.\\
Single Hop Source Routing (SOSR) \cite{gummadi2004improving} & 2004	 & Uses randomly chosen intermediaries to route packets indirectly towards the eventual destinations in order to recover from failures which occur near the destinations. \\
SPREAD \cite{lou2004spread} & 2004	 & Divides a secret message into multiple parts and sends them over multiple independent paths towards the destination in order to improve secrecy and security. \\
Routing Deflections \cite{yang2006source} & 2006 & Provides path diversity by allowing routers to redirect packets based on tagged information provided by end hosts, subject to the compliant of the ISP routing policy.  \\
Joint Multi-channel Multi-path Control (JMM)  \cite{tam2007joint} & 2007 & Achieves a balanced distribution of the contending input traffic at different times over different channels and different paths through cross-layer interaction between link and network layers. \\
New Internet Routing Architecture (NIRA) \cite{yang2007nira} & 2007 & Provides path diversity by enabling a packet, in a source controlled manner, to switch among multiple ISPs enroute to its destination.   \\
Path Splicing \cite{motiwala2008path} & 2008 & Constructs a path by combining slices of multiple routing trees, which serve as backups paths, en route to an eventual destination to improve reliability. \\
Pathlet Routing \cite{godfrey2009pathlet} & 2009 & Performs routing deflection at the domain level of a hierarchical network. Each domain advertises a few paths or \textit{pathlets} used by the a source from another domain. \\
Slick Packets \cite{nguyen2011slick} & 2011	&  Embeds alternate route information is embedded in the packet's header in a source controlled fashion to avoid failures. This scheme requires minimal network routing state because of the source routing feature. \\
AMIR \cite{qin2012amir} & 2012	& Establishes multiple paths around the primary paths in hierarchical networks. \\

\\
\multicolumn{2}{l}{\textbf{\underline{\emph{Hop-by-Hop Routing}}}}\\
\\

Dispersity routing \cite{maxemchuk1975} & 1975 & Sends distinct data packets over multiple paths for load balancing,
or sends duplicated data packets over multiple paths for improved reliability and resiliency.  \\
Geographical Power Efficient Routing (GPER) \cite{wu2004gper} & 2004 & Uses probabilistic multipathing to achieve path diversity. The nodes are rendered to choose their transmission ranges to conserve energy. \\
Loop-free Invariant (LFI)-based Approaches \cite{vutukury1999simple}, \cite{vutukury1999algorithm}, \cite{vutukury2001mdva} & 1999, 2001 & Incorporates multipath routing in traditional link state-based and distance vector-based algorithms to find loop-free multiple paths using LFI condition. \textit{Diffusing computation} is used in \cite{vutukury2001mdva}. \\
Hashing-based ECMP \cite{cao2000performance} & 2000 & Selects paths from a set of \textit{equal cost multiple paths} in order to split flows for load balancing. It comprises of direct hashing and table-based hashing methods. \\
Srinivas and Modiano \cite{srinivas2003minimum} & 2003 & Uses 2-link-disjoint and k-node-disjoint algorithms to conserve energy. \\
Packet-caching \cite{valera2003cooperative}, \cite{zhang2010reliable} & 2003, 2010 & Circumvents route breakdowns by caching/ storing (a fixed number of) data packets along with alternate route information at nodes in a distributive fashion. \\
Interference Minimization Approaches (M-MPR \cite{de2003meshed, de2003WCNC};  Maimour \cite{maimour2008maximally}) & 2003, 2008 & Constructs multiple paths in an incremental fashion keeping in view the congestion and interference effects (using the interpath interference and disjointness metrics). M-MPR incorporates FEC and selective forwarding. \\
AODV-enhanced Routing Schemes (AOMDV \cite{marina2001demand}; Ye et al.  \cite{ye2003framework}; MP-AODV \cite{sambasivam2004dynamically}; NDMR \cite{li2004demand}) & 2001, 2003, 2004, 2004 & AOMDV establishes multiple link-disjoint paths in an on-demand manner; Ye et al. use reliable nodes, in terms of energy resources, to establish multiple paths; MP-AODV selects primary paths using received signal strength, and alternative paths through heartbeat message exchanges; NDMR incorporates mutipathing into traditional AODV and DSR, and uses alternative paths as backups. \\
Biologically Inspired Routing \cite{di2005anthocnet, ducatelle2005ant} & 2005 & Uses biologically inspired approaches (e.g., ant colony optimization) to establish and maintain multiple paths. \\
Security-Enhanced Approaches (SEIF, SMBR, H-SPREAD) & 2006, 2011 & Enables nodes to take independent decisions to address joint fault and intrusion tolerance. \\
Enhanced BGP \cite{kushman2007r}, \cite{van2009loop} & 2007, 2009 &  Extends the traditional BGP routing protocol to establish and maintain loop-free multiple paths in the face of multiple failures in hierarchical networks. \\
Maximum Alternative Routing Algorithm (MARA) \cite{ohara2009mara} & 2009 & Constructs directed acyclic graphs (DAGs) that (i) maximize the min. connectivity; (ii) maximize the min. max. flows; and (iii) maximize the min. max. flows as an extension of shortest-path routing.\\
Huang and Fang \cite{huang2009performance} & 2009 & Selects multiple parent nodes in an opportunistic manner to establish multiple paths even though nodes may have unknown speeds in vehicular ad-hoc networks. \\
Network coding based reliable disjoint and braided multipath routing protocol \cite{yang2010network} & 2010 & Uses braided and disjoint multiple paths in a distributed fashion in conjunction with network coding for reliability and energy conservation purposes. \\


Yet another multipath routing (YAMR) \cite{ganichev2010yamr} & 2010 &  Establishes multiple paths, and provides concurrent transmission with minimal overhead in hierarchical networks. \\
Beltagy et al. \cite{beltagy2011new} & 2011 & Uses multiple paths to reduce negative effects from and to licensed users in cognitive radio networks.\\

\\
\multicolumn{2}{l}{\textbf{\underline{\emph{Overlay Routing}}}}\\
\\

Deroute \cite{savage1999end} & 1999 & Proposes an edge-based routing scheme for the Detour overlay network comprising geographically distributed routes for better end-to-end performance. \\

Resilient Overlay Network (RON) \cite{andersen2001case} & 2001 & Focuses on equipping end hosts and applications with the ability to switch between direct routing (through the underlying network) or indirect routing (using the overlay network). \\
Multipath Video Streaming on Overlay Networks \cite{ma2004new} & 2004 & Proposes a link correlation-based QoS metric along with path correlation model for multipathing using an overlay network for efficient video streaming.\\
Multipath multiple description video streaming \cite{begen2005multi} & 2005	&  Selects a set of multiple paths based on \textit{multiple description coding} models.\\
BARON \cite{lee2008bandwidth} & 2008 & Proposes a bandwidth-aware routing scheme for overlay networks to meet application bandwidth requirements and improve the overall throughput performance.\\

\\

\multicolumn{2}{l}{\textbf{\underline{\emph{SDN-Based}}}}\\
\\

Hedera \cite{al2010hedera} & 2010 & Proposes an efficient, scalable, adaptive and central scheduling method, with a universal scope of active flows in view, for multistage switching processes in datacenters.\\
Micro-TE \cite{benson2011microte} & 2011 & Implements central controllers using OpenFlow, as well as uses short-term traffic predictability and multipath routing, to schedule flows in a datacenter environment. \\
Rain Man \cite{stephens2012designing} & 2012 & Proposes a layer two (L2) multipath packet forwarding algorithm for a SDN-based firmware architecture for datacenters.\\
B4 (Google) \cite{jain2013b4} & 2013 & Proposes a centralized TE solution that utilizes OpenFlow/ SDN principles along with multipath forwarding to improve network efficiency (i.e., link utilization and average network utilization). \\
Wireless Mesh Software-defined Networks (wmSDN) \cite{detti2013wireless} & 2013 & Achieves traffic balancing and selects gateways in an intelligent round robin fashion for Internet connectivity. \\
Software Defined VANETs \cite{kutowards} & 2014 & Establishes multiple paths to address route failures, as well as for the transmission of high priority flows, in VANETs. \\

\\
\hline
\end{tabular}
\end{table*}

\begin{table*}
\scriptsize
\ssmall
\caption{Summary of General Multipath Routing Proposals (Designed Mainly For \textit{Wired Networks})}
\centering
\label{tab:MultipathRouting}
\begin{tabular}{p{2.5cm}p{0.5cm}p{1.7cm}p{1cm}p{2.3cm}p{7.7cm}}

\toprule
\cellcolor[HTML]{EFEFEF}\textbf{\textit{Work}} & \cellcolor[HTML]{EFEFEF}\textbf{\textit{Year}} & \cellcolor[HTML]{EFEFEF}\textbf{\textit{Route Construction}} & \cellcolor[HTML]{EFEFEF}\textbf{\textit{Paths}} & \cellcolor[HTML]{EFEFEF}\textbf{\textit{Metric(s)}} & \cellcolor[HTML]{EFEFEF}\textbf{\textit{Description}}\\
\midrule

Dispersity Routing \cite{maxemchuk1975} & 1975 & Hop-by-Hop & Concurrent & Delay, Reliability, Load Balancing, Link Utilization & Disperses traffic over multiple paths either in a redundant or non-redundant manner. \\
OSPF-OMP \cite{villamizar1999ospf, villamizar1999mpls} & 1999 & Hop-by-Hop & Concurrent & Delay, Load Balancing & Provides a hashing-based OSPF extension to support optimized multipathing for load balancing. \\
ECMP \cite{hopps2000analysis} & 2000 & Hop-by-Hop & Concurrent & Delay, Load Balancing & Forwards packets based on a CRC16 hash to one of the nodes forming equal-cost paths towards a destination. \\

Multipath MPLS \cite{seok2001dynamic} & 2001 & Source Routing & Concurrent & Link Utilization, \newline Load Balancing & Constructs multiple paths in polynomial time subject to a constraint (i.e., number of hops and paths). \\

Feedback-based Routing (FBR)  \cite{zhu2003feedback} & 2003 & Source Routing & Backup & Delay & Enables an access router to compute two routes for each network prefix based on the structural information disseminated by transit routers.\\

BANANAS \cite{kaur2003bananas} & 2003 & Source Routing & Concurrent & Resource Utilization & Provides a framework that creates \emph{global identifier} to enable multipathing in the Internet. \\
SOSR \cite{gummadi2004improving} & 2004 & Source Routing & Backup & Overhead, Reliability & Presents a \emph{single hop source routing (SOSR)} to avoid failures which occur near the destinations in the Internet. \\
MONET \cite{andersen2005improving} & 2005 & Overlay Routing & Backup & Overhead, Delay, \newline Reliability & Provides multiple backup paths through proxies to websites for enhanced web availability. \\
Routing Deflections \cite{yang2006source} & 2006 & Source Routing & Concurrent & Overhead, Reliability & Provides an ISP that constructs alternate paths or \emph{deflections} based on the \emph{tags} that users put in their packets. \\
MIRO \cite{xu2006miro} & 2006 & Interdomain Negotiations & Backup & Reliability & Provides interdomain negotiations while establishing paths by improving upon BGP and source routing. \\
R-BGP \cite{kushman2007r} & 2007 & Hop-by-Hop & Backup & Reliability & Provides alternate paths to overcome transient failures by modifying BGP. \\
NIRA \cite{yang2007nira} & 2007 & Source Routing & Backup & Reliability, Overhead & Enables a user to choose the sequence of ISPs for its packets in order to switch among multiple routes. \\
Path Splicing \cite{motiwala2008path} & 2008 & Source Routing & Backup & Reliability & Allows \emph{slices} of different routing trees to be spliced together for reliable delivery of a packet towards a destination. \\
D-BGP and B-BGP \cite{wang2009path} & 2009 & Hop-by-Hop & Backup & Overhead, Reliability & Provides a path diversity-aware routing protocols in which an advertisement method is proposed for multiple paths. \\
Pathlet Routing \cite{godfrey2009pathlet} & 2009 & Source Routing & Backup & Reliability & Allows a source to concatenate and advertise segments of paths or \emph{pathlets} in packets while constructing an end-to-end path. \\
Multipath BGP \cite{van2009loop} & 2009 & Hop-by-Hop & Concurrent & Resource Utilization,  & Uses multiple paths in a concurrent fashion and preserves loop-freeness by modifying BGP. \\
YAMR \cite{ganichev2010yamr} & 2010 & Hop-by-Hop & Backup & Reliability, Overhead & Constructs a set of multiple paths to avoid (at least) an interdomain link failure. \\


Multi-path BGP \cite{valera2011multi} & 2011 & Hop-by-Hop & Concurrent & Resource Utilization & Proposes two multipathing approaches at the interdomain level. \\
AMIR \cite{qin2012amir} & 2012 & Source Routing & Concurrent & Reliability & Constructs alternative paths around the primary path using topology information from neighboring ASes. \\

\bottomrule
\end{tabular}
\label{tab:MultipathRouting}
\end{table*}

\begin{table*}
\scriptsize
\ssmall
\caption{Summary of Multipath Routing Proposals for \textit{Wireless Networks}}
\centering
\begin{tabular}{p{2.7cm}p{0.3cm}p{1.58cm}p{.8cm}p{2.1cm}p{7.6cm}}

\toprule
\cellcolor[HTML]{EFEFEF}\textbf{\textit{Work}} & \cellcolor[HTML]{EFEFEF}\textbf{\textit{Year}} & \cellcolor[HTML]{EFEFEF}\textbf{\textit{Route Construction}} & \cellcolor[HTML]{EFEFEF}\textbf{\textit{Paths}} & \cellcolor[HTML]{EFEFEF}\textbf{\textit{Metric(s)}} & \cellcolor[HTML]{EFEFEF}\textbf{\textit{Description}}\\
\midrule
\\
\textbf{\textit{\underline{Wireless Sensor Networks (WSNs)}}}\\

Disjoint and Braided \cite{ganesan2001highly} & 2001 & Hop-by-Hop & Backup & Energy, Reliability, \newline Overhead & Proposes a reliable, non-disjoint or \emph{braided multipathing} scheme that performs 50\% better than disjoint multipathing. \\
Selective Forwarding \cite{de2003meshed, de2003WCNC} & 2003 & Hop-by-Hop & Concurrent & Energy, Throughput & Proposes a meshed multipath routing protocol with selective forwarding. \\

MMSPEED \cite{felemban2006mmspeed} & 2006 & Overlay Routing & Concurrent & Delay, Reliability & Proposes a probabilistic multipath selection scheme under reliability and QoS constraints (i.e., delay). \\
H-SPREAD \cite{lou2006h} & 2006 & Source Routing & Concurrent & Reliability & Proposes a hybrid multipath method where node/ link-disjoint paths are maintained on each sensor node. \\
EBMR \cite{chen2006energy} & 2006 & Deterministic\newline(Defined by BS) & Backup & Energy & Enables a BS to discover, maintain and switch among multiple paths keeping in view the nodal residual energy. \\

Totally Disjoint Paths \cite{waharte2006totally} & 2006 & Hop-by-Hop & Concurrent & Interference, Throughput & Proposes and studies a 2-path routing scheme in a single and a multiple source-destination pair scenarios. \\
 GPER \cite{wu2007power} & 2007 & Hop-by-Hop & Backup & Energy & Enables nodes to decide their transmission ranges distributively for power efficiency using probabilistic multipathing. \\
 DMPR \cite{ramasubramanian2007disjoint} & 2007 & Hop-by-Hop & Concurrent & Load Balancing, \newline Throughput & Constructs two (colored) trees, which are rooted at the sink node. Two paths are formed from a source via each tree, which is link/ node disjoint. \\
SEEM \cite{nasser2007seem} & 2007 & Deterministic\newline(Defined by BS) & Backup & Energy, Throughput, \newline Overhead, Reliability & Enables BS to discover and maintain multiple routes, as well as switch among them, in order to conserve energy and enhance security. \\
Multicast Multipath \cite{wu2007qos} & 2007 & Hop-by-Hop & Concurrent & Delay, Bandwidth & Proposes three tree-based on-demand and distributed multipath routing methods under multicast QoS constraints. \\
Ming et al. \cite{ming2007energy} & 2007 & Hop-by-Hop & Concurrent & Energy, Delay, Overhead & Presents a scalable, distributed, reliable and node-disjoint multipath routing protocol to achieve efficient load balancing. \\
Maximally Radio Disjoint Routing \cite{maimour2008maximally} & 2008 & Hop-by-Hop & Concurrent & Energy, Bandwidth, Interference & Constructs multiple paths in an incremental way to reduce interference in multimedia WSNs. \\
Multiconstrained QoS Routing \cite{huang2008multiconstrained} & 2008 & Hop-by-Hop & Concurrent & Delay, Reliability & Constructs multiple paths based on a probabilistic link-state method under delay and reliability QoS constraints. \\
REER \cite{yahya2009reer} & 2009 & Hop-by-Hop & Both & Energy, Delay, Reliability & Uses backup and concurrent multipathing approaches to conserve energy and enhance robustness. \\
NC-RMR \cite{yang2010network} & 2010 & Hop-by-Hop & Concurrent & Energy, Reliability & Uses both disjoint and braided multiple paths, with network coding, in a per-hop fashion for energy conservation and reliability enhancement. \\
SEIF \cite{challal2011secure} & 2011 & Hop-by-Hop & Concurrent & Energy, Reliability & Proposes a distributed, BS
independent and one-way hash chains for multipathing. \\
IM2PR \cite{radi2014im2pr} & 2014 & Hop-by-Hop & Concurrent & Energy, Reliability & Presents an inteference-minimized multipath routing protocol (IM2RP) that aims to discover minimum interfering paths.\\

\\

\textbf{\textit{\underline{Mobile Ad-hoc Networks (MANETs)}}}\\
On-demand \cite{nasipuri1999demand} & 1999 & Source Routing & Backup & Overhead & Modifies DSR to support multipathing. \\
MSR \cite{wang2000multipath} & 2000 & Source Routing & Concurrent & Load Balancing & Uses a probing-based mechanism to provide multipath support in DSR for load balancing. \\
SMR \cite{lee2001split} & 2001 & Source Routing & Concurrent & Delay, Overhead, \newline Load Balancing & Constructs two maximally disjoint paths, with one being the smallest delay path, in an on-demand fashion. \\
AOMDV \cite{marina2001demand} & 2001 & Hop-by-Hop & Backup & Reliability & Modifies AODV to construct multiple link-disjoint paths using the advertised hop count information. \\
MP-DSR \cite{leung2001mp} & 2001 & Source Routing & Concurrent & Bandwidth, Reliability & Provides a QoS-aware multi-path dynamic source routing protocol for MANETs. \\
Ye et al. \cite{ye2003framework} & 2003 & Hop-by-Hop & Concurrent & Reliability & Modifies AODV to support multiple node-disjoint paths by introducing a few \emph{controlled reliable nodes} in the topology. \\

CHAMP \cite{valera2003cooperative} & 2003 & Hop-by-Hop & Concurrent & Reliability, \newline Resource Utilization & Circumvent frequent link breakdowns by caching (a few) data packets along with alternate route information at nodes distributively. \\

Min. Energy Paths \cite{srinivas2003minimum} & 2003 & Hop-by-Hop & Concurrent & Energy & Provides two algorithms to construct 2-link-disjoint and k-node-disjoint paths to conserve energy in dense and sparse networks, respectively. \\
SPREAD \cite{lou2004spread} & 2004 & Source Routing & Concurrent & Reliability & Sends data in the form of \emph{shares} over multiple independent paths to ensure confidentiality. \\

Serial Multiple Disjoint Trees Multicast Routing (Serial MDTMR) \cite{wei2004multipath} & 2004 & Source Routing & Concurrent & Robustness & Provides real-time efficient video streaming using multiple paths for unicast and multicast traffics. \\

MP-AODV \cite{sambasivam2004dynamically} & 2004 & Hop-by-Hop & Backup & Reliability & Provides adaptive multipathing in AODV based on signal strength received at nodes. \\
NDMR \cite{li2004demand} & 2004 & Hop-by-Hop & Backup & Delay, Reliability & Constructs node-disjoint multiple paths for a source-destination pair. \\

Chen et al. \cite{chen2005multipath} & 2005 & Hop-by-Hop & Concurrent & Delay, Reliability, \newline Load Balancing & Presents an analytical model for multipath routing in MANETs for optimal load balancing, as well as to address delay and reliability issues. \\

 Multipath OLSR \cite{kun2005research} & 2005 & Hop-by-Hop & Concurrent & Energy, Load Balancing & Modifies OLSR to support multipathing. \\
 AntHocNet \cite{di2005anthocnet, ducatelle2005ant} & 2005 & Hop-by-Hop & Concurrent & Delay, Reliability, Jitter & Provides a hybrid multipathing algorithm with reactive path setup, as well as proactive path probing and maintenance, based on an \emph{ant-colony optimization framework}. \\
  SDMR \cite{galvez2008spatially} & 2008 & Source Routing & Concurrent & Load balancing, Bandwidth, Interference, Path Closeness & Enables a source node to calculate multiple spatially disjoint paths based the information extracted from route reply packets from the destination node. \\
  
Multipath Routing Reliability \cite{caleffi2008reliability} & 2008 & Overlay Routing & Backup & Reliability & Analyzes the reliability of multipath routing protocols based on an overlay graph concept framework. \\

RAMP \cite{zhang2010reliable} & 2010 & Hop-by-Hop & Concurrent & Delay & Jointly addresses the issues of multipath routing and deferential delay. \\

Mobile Cloud \cite{kusoftware} & 2014 & SDN-based & Concurrent & Reliability & Uses a central controller in a SDN fashion to perform multipath routing in MANET. \\
\\
\textbf{\textit{\underline{Wireless Mesh Networks (WMNs)}}}\\

Weighted Interference Multipath (WIM) \cite{tsai2007interference} & 2007 & Source Routing & Concurrent & Interference, Delay, \newline Reliability & Selects multipath based on \emph{weighted mutual-interference (WIM)} metric in a multi-channel multi-radio environment. \\

Joint Multichannel Multipath (JMM) \cite{tam2007joint} & 2007 & Source Routing & Concurrent & Throughput & Splits traffic among different channels at different times over multiple paths. \\

wmSDN \cite{detti2013wireless} & 2013 & SDN-based & Backup & Load Balancing, \newline Resource Utilization & Selects Internet gateways of a mesh network in a round robin fashion to balance outgoing traffics using a central controller. \\

\textbf{\textit{\underline{Cognitive Radio Networks (CRNs)}}}\\

Beltagy et al. \cite{beltagy2011new} & 2011 & Hop-by-Hop & Backup & Reliability, Path Closeness & Constructs multiple routes to minimize the effects of PU activity based on a \emph{closeness} metric. \\

Ali et al. \cite{ali2015} & 2015 & Hop-by-Hop & Backup & Reliability & Proposed a learning automata based multipathing solution for multicast routing in CRNs. \\

\textbf{\textit{\underline{Vehicular Ad-hoc Networks (VANETs)}}}\\

Huang et al. \cite{huang2009performance} & 2009 & Hop-by-Hop & Concurrent & Delay, Interference, \newline Reliability & Constructs node-disjoint multiple paths based on the mutual interference of paths. \\

LIAITHON \cite{wang2012liaithon} & 2012 & Hop-by-Hop & Concurrent & Load Balancing, Interference, Path Closeness & Provides a receiver-initiated video streaming multipath solution in which nodes comprising the multiple paths are chosen so that they have minimum coupling/closeness.\\

Software defined VANETs \cite{kutowards} & 2014 & SDN-based & Backup & Reliability & Deploys VANETs in a software-defined environment for reliability enhancement and power management using a central controller. \\

\textbf{\textit{\underline{Personal Area Networks (PANs)}}}\\

IEEE 802.15.4 \cite{pavkovic2011multipath} & 2011 & Hop-by-Hop & Concurrent & Delay, Reliability & Chooses a parent node opportunistically to enable multipathing by combining IEEE 802.15.4's MAC layer \emph{cluster-tree} method with RPL's \emph{destination-oriented directed acyclic graph (DODAG)}.\\

\bottomrule
\end{tabular}
\label{tab:MultipathRoutingWireless}
\end{table*}

\subsection{Source Routing}
\label{subsec:source_routing}

In many cases, the network is not well placed to know about the optimality of a route. The source host/ user, on the other hand, knows about the application's unique requirements and can judge about the appropriateness of a route. The source is also best suited to deal with any failures that may arise along a route. This motivates \textit{source routing} in which the source host, or edge router, defines the end-to-end path to be used for forwarding packets with the potential to improve network performance, reliability and user satisfaction.  

In the source routing paradigm, the source explicitly embeds information in a data packet about the path that the packet will traverse before reaching its destination \cite{yang2006source}. Such a scheme proceeds typically by first discovering routes through the exchange of control messages and then embedding the discovered route information in the data packet's header. The routers along the explicitly defined path then forward the data packet using the information stored in the data packet's header. Although source routing provides maximum flexibility, it is important to point out some difficult challenges that it faces \cite{xu2006miro} including (i) limited control of intermediate autonomous system (AS); (ii) scalability problems due to the need of having the source to compute and define routes; and (iii) efficiency and stability problems that may arise from selfish source routing.

In the next two subsections, we will describe source routing proposals, initially discussing general solutions (developed mainly for wired networks) in Section \ref{subsub: w_g_source_routing}, and then discussing source routed multipath solutions specific to wireless networking in Section \ref{subsub:wireless_source_routing}.

\vspace{2mm}
\subsubsection{Source Routed Multipathing (General/ Wired Solutions)}
\label{subsub: w_g_source_routing}



To accompany this section, we present a comprehensive listing of general multipathing solutions (designed primarily for wired networks) in Table \ref{tab:MultipathRouting}. This table serves as a convenient summary to complement the description of various hop-by-hop multipating routing solutions in this section. In addition, the table also includes information about works that have used other routing schemes (source routing, overlay routing, and SDN-based routing), and we will be revisiting this table later in this paper. We note here that the classification is approximate since not all works can be cleanly pigeonholed into only one of our broadly defined categories.

Due to the lack of flexible support for traffic engineering in IP-based networks, the multiprotocol label switching (MPLS), with its rich support for traffic engineering (TE), has become ubiquitously deployed in ISPs. A lot of work has been done on efficient optimized multipath TE. Lee et al. \cite{lee2002constrained} proposed a constrained multipath TE scheme for MPLS-based networks to help find optimal paths that meet the provided traffic demand as well as the constraints (such as maximum hop count and node/ link preferences). The proposed scheme calculated the flow split ratio for the multiple paths. The evaluation of the proposed scheme showed superior results in utilization, traffic volume, and overhead compared to the conventional shortest path routing scheme. In another work, Seok et al. \cite{seok2001dynamic} proposed to integrate MPLS traffic engineering with source multipath routing subject to a constraint on the total number of hops and the total number of paths. The presented algorithm calculated the flow split ratio by using hashing over the calculated multiple paths as per the traffic demands dictated by an ingress-egress pair of a network. A dynamic traffic engineering approach is proposed to tackle the issue of congestion as the heuristic presented in this work protects a link from the overwhelming effect of huge traffic demands by minimizing the maximum utilization of a link and using the aspect of multipathing for load balancing.

\vspace{2mm}
\textit{Approaches to Avoid Failure/ Performance Bottlenecks}
\vspace{2mm}

It was shown in \cite{savage1999end} that upto 80\% of the default routing paths on the Internet are suboptimal with an alternative route offering potentially lower loss rate. This has motivated the development of numerous \textit{source-routed} routing protocols that can help route around the performance bottleneck/ failed entity.

Gummadi et al. \cite{gummadi2004improving} proposed a source routing-based approach named ``\textit{scalable one-hop source routing}'' (SOSR) to recover from Internet path failures by routing indirectly through randomly chosen intermediaries. It has been shown in the previous work that last-hop and end-host failures account for a disproportionately high number of path failures. Such failures are especially hard to recover as there is a limited number of network elements to route through as the route nears the destination. SOSR recovers from such failures by routing packets indirectly using a small set of \textit{randomly chosen} network intermediaries. When a network failure occurs, then the source probes the destination through the randomly chosen set of intermediary nodes. If such a probe is successful, then the data is routed via this indirect route; otherwise, it is concluded that the destination is now unreachable. In this technique, since the choice of the intermediary router is made randomly from a preselected set, the continuous maintenance of backup routes is avoided. This allows SOSR to outperform other overlay or backup path techniques that require continuous path maintenance and are more expensive in terms of control overhead.

Using the concept of \emph{path splicing} \cite{motiwala2008path}---which can be applied in both interdomain and intradomain routing scenarios---a packet can traverse across segments or \textit{slices} of different routing trees (based on the discretion of the end hosts). In path splicing, multiple routing trees are established with a packet able to switch from one routing tree to another at any hop/ edge. This concept has been shown to improve network resilience.

\vspace{2mm}
\textit{Interdomain Source-Routed Multipath Routing}
\vspace{2mm}

To manage a global-scale routing, the approach adopted on the Internet is to separate routing into intradomain routing and interdomain routing, where the former focuses on ``optimal'' routing (defined in the terms of some optimization metrics) within a single AS, while the latter focuses on ensuring loop-free policy compliant routing between multiple ASes. Since most traffic on the Internet traverses across multiple domains, interdomain multipathing is of great practical importance. The defacto interdomain routing protocol of the Internet is \textit{Border Gateway Protocol (BGP)}. BGP allows different ASes to exchange reachability information that is then used by BGP for computing loop-free policy-compliant routing. 

Although BGP is well entrenched in the Internet ecosystem, it is widely known that most routes chosen by BGP are not optimal in terms of performance, cost, or reliability. There are many challenges involved in integrating source routing, multipathing, and global Internet routing \cite{he2008toward}. A major impediment being the rigid inflexible policies of various ASes for interdomain routing \cite{he2008toward}---e.g., it is typical for BGP to ignore potential paths suggested by a source routing scheme \cite{he2008toward}---which makes the control of the end-to-end routing behavior difficult for end hosts or even any single AS. There are ongoing efforts to devise new techniques to adapt BGP to implement source routing while supporting the provisioning of multiple paths through ASes \cite{kushman2007r, wang2009path, van2009loop}. With various trends on the Internet (such as multihoming) nudging the Internet towards multipathing, there is also an increased interest in adapting interdomain routing protocols for multipath. Different works have attempted to develop new techniques based on the legacy interdomain routing protocols that we describe next. 

Feedback-based routing (FBR) \cite{zhu2003feedback} is an alternative approach to interdomain routing. In contrast to BGP, this approach differentiates routing information into its structural components (that deals with the existence of links) and dynamic components (which refers to the quality of these links). In this approach, the core routers of the Internet are only responsible for disseminating structural information, while the routing decisions are taken at the edge based on dynamic end-to-end performance measurements in addition to the structural information. FBR is multipath-based since each access router computes two paths to every advertised network prefix (that differs from each other as much as possible)  based on the structural information it receives. The access router then monitors the route quality, and uses the better path to carry its traffic. In case the access router finds that both the routes have an infinite RTT, then two new routes are computed. Since the core/ transit routers  do not compute routes (and only disseminate structural information), FBR scales better than BGP.

The caveat of source routing is the overhead involved in carrying the routing information in the packet itself. Various approaches have been proposed for this. LIPSIN proposes putting a \textit{Bloom filter} into data packets in source routing style as a forwarding fabric for multicast traffic for publish/ subscribe Internets \cite{jokela2009lipsin}. MPLS can be used for explicit routing, which is similar to routing, by using labels avoiding the overhead of carrying the information of the entire path in the packets. The \textit{BANANAS} architectural framework \cite{kaur2003bananas}, on the other hand, works by encoding paths using a short hash (called the PathID) which is a sequence of globally-known identifiers. 

In the BANANAS framework \cite{kaur2003bananas}---which can be deployed in both intradomain or interdomain settings----is an attempt to make  conventional routing protocols (such as OSPF, ISIS and BGP) capable of exploiting the path diversity of the Internet. The need of specifically constructing multiple paths is eliminated by generating a short hash called a \textit{PathID} using a globally accessible sequence of identifiers. A path is encoded as PathID using a hash on a sequence of identifiers (or any non-null permutation of such sequence) that can be the IDs of routers, link interfaces, link weights, AS numbers (used in the case of interdomain routing), etc.  The global significance of these identifiers distinguishes BANANAS from the conventional MPLS and ATM architectures. Routers that are multipath-enabled can calculate a PathID with the global scope of the above-mentioned identifiers in view. This feature makes the routers independent and autonomous with respect to the setup of multiple paths without the need of \textit{explicit} control signaling protocol for a specific path setup. 


Multipath Interdomain Routing (MIRO) \cite{xu2006miro} is a BGP-based interdomain multipath routing framework that is backward compatible with BGP and incrementally deployable. In MIRO, the default routes are learnt through the BGP protocol, and arbitrary pairs of domains can perform AS-level negotiations to add additional paths tailored to their specific needs. MIRO provides more flexible control of traffic to the intermediate ASes with reasonable overhead. This is done to promote path diversity and improve path resilience in interdomain routing. The main features of MIRO include (i) AS-level path selection; (ii) negotiation support for alternate routes between two arbitrary ASes (that may not be adjacent); and (iii) support for policy-driven export of alternate routes by the responding AS during negotiation (through which transit AS have greater control over which traffic enters their networks). Although MIRO is similar to overlay routing in that it uses tunnels, it differs from overlay schemes because it allows paths to be selected in the underlay through the cooperation of intermediate ASes. The difference between MIRO and most of the other source routing works is that, apart from providing end hosts or the edge routers with routing control, MIRO allows control over path selection by intermediate ASes. 

The ``new Internet routing architecture'' (NIRA) \cite{yang2007nira} is a source-routed interdomain multipath routing protocol in which the end host has more flexible control over the sequence of ASes to be traversed by its packets. By providing the end-host control over the adopted end-to-end path, NIRA can help foster healthy ISP competition leading to enhanced services and improved end-to-end performance and reliability. NIRA provides users/ end hosts with more control over the chosen routes without running a global link-state (LS) routing protocol, and thus NIRA can provide a user with a route with lower control overhead. With NIRA, a user only specifies a source address and a destination address to choose a specific interdomain path comprising multiple ASes, and the path is altered when the source or the destination addresses is changed. This technique is efficient because of its faster fail-over action and lower control overhead.

The concept of routing deflections \cite{yang2006source} is applied at the level of domains in \textit{pathlet routing} \cite{godfrey2009pathlet}. In pathlet routing, domains advertise multiple paths or \textit{pathlets} to a source located in another domain. These pathlets afford the source with the ability to exploit the advertised path diversity through route deflections. More abstractly, pathlet routing allows source routing over a virtual topology. Using the concept of routing over virtualized topologies, network operators can declare their routing services expressively in a flexible fashion. In particular, using pathlet routing, the routing styles of BGP, source routing, MIRO, LISP and NIRA can be emulated \cite{godfrey2009pathlet}.

AMIR is a multipath interdomain routing protocol proposed in \cite{qin2012amir} that establishes multiple paths and deploys source routing at AS level. In this scheme, the multiple paths are established around a primary path that is usually constructed by a policy-compliant traditional interdomain routing protocol. This is achieved by negotiating path provisioning with neighboring ASes. The participating ASes are called \textit{multipath service agents (MPSA)} and these agents impose charges for providing topological information for the computation of multiple paths through them. Data is routed over these ASes by a \textit{multipath service forwarding agent (MSFA)} embedded at the edge routers of such ASes. Paths are set up and the data is forwarded in a source routing fashion using MSFAs of MPSAs. AMIR enhances application-level user experience with lower overhead incurred in path setup at the AS level.

\vspace{2mm}
\subsubsection{Source Routed Multipathing (Wireless Solutions)}
\label{subsub:wireless_source_routing}

There are various source-routed multipath routing proposals for wireless networks---with the major bulk of work being done for mobile ad-hoc networks (MANETs) and wireless mesh network (WMNs). In this subsection, we will cover the wireless networking configurations of wireless sensor networks (WSNs), MANETs, and WMNs.

\vspace{2mm}
\textit{Source-Routed Multipath Protocols for WSNs:}
\vspace{1mm}

WSN multipath routing protocols have rarely adopted source-routed solutions to avoid the scalability problem that may arise by overburdening 
the potentially energy constrained source node. The major bulk of multipath protocols proposed for WSNs are based on the hop-by-hop routing paradigm instead that will be covered in Section \ref{subsec:hop-by-hop}. As an example of a source-routed WSN multipath solution, we will discuss the H-SPREAD \cite{lou2006h} routing protocol that facilitates the construction of multiple node-disjoint paths towards the base station in a distributed manner. In such a setting, only a few paths can be compromised if some malicious nodes appear in the network. If the next-hop node of a packet at an intermediate node is unavailable, then this node redirects the packet (in a source routing manner) to an alternate next hop which is part of an available backup path, thus enhancing the reliability of a data transfer. It is only when there is no available alternate path that a packet is dropped at a node.

\vspace{2mm}
\textit{Source-Routed Multipath Protocols for MANETs:}
\vspace{1mm}

Most of the routing protocols proposed for MANETs are reactive/ on-demand. On-demand routing protocols compute routes when they are needed for communication and are better suited to dynamic topologies such as MANETs. Such an approach also avoids the extra overhead associated with proactive routing (due to the periodic exchange of control packets). In on-demand routing, routes are discovered through the flooding of specific control packets by the source that are used to gather the topological information. This information is then collected using a reply packet from the destination. The legacy protocol for MANETs called  \textit{dynamic source routing (DSR)} \cite{johnson1996dynamic} works on the same principle. After computing the route, a source node embeds the route information in a packet's header. These packets are then forwarded by the intermediate mobile nodes to the eventual destination.

In \cite{nasipuri1999demand}, Nasipuri et al. have modified DSR so that it can support the provisioning of disjoint multiple paths. Two flavors of multipath routing are presented: in the first approach, only the source node is provisioned with multiple paths; while in the other, both source and intermediate nodes (that make up the primary path in the normal operation of DSR) are also provided with multiple disjoint paths around the primary path. This work makes an effort to reduce the control overhead incurred by the conventional DSR in the route discovery process. This control overhead is a consequence of flooding the query packets in the network. The reduction in the control overhead is due to the provisioning of multiple backup paths. So, in the case when the primary path fails, the traffic is switched to one of the backup paths without rediscovering the route all over again.

Another similar work based on modifications in DSR is called \textit{multipath source routing} (MSR) that is proposed in \cite{wang2000multipath}. Multiple paths are computed using the conventional DSR's route discovery and maintenance operations. In this work, a probing-based mechanism is used to achieve a balanced load over multiple constructed paths. This mechanism probes multiple paths for their RTT responses and the traffic is load balanced accordingly to control congestion and reduce end-to-end delay.

Two maximally disjoint paths are created to balance the input traffic over them. One of the paths constructed is the conventional shortest path. This process is done in an \textit{on-demand} fashion. This work is called \textit{split multipath routing} (SMR) \cite{lee2001split} and it performs better than DSR in terms of end-to-end delay. Traffic is distributed over the multiple paths on a per-packet manner. This technique, although effective for load balancing and controlling congestion, can create a packet-reordering problem at the destination that makes its use with TCP problematic.

Multiple independent paths are deployed for the purpose of improving secrecy and security in \cite{lou2004spread}. This method is called \textit{SPREAD}. In this method, a secret message is divided into multiple parts called \textit{shares}. Then, using secret (threshold) sharing methods, these shares are routed over multiple independent paths towards the destination. An adversary node has to compromise at least a certain number of shares (a specific threshold number) out of the total to completely decipher the original message. To enhance the security further, the shares are forwarded over multiple independent paths towards a destination. In this case, an attack on all the paths must be made by salvaging at least the threshold number of shares to decode the original message. By this way, even if some of the nodes along the paths are compromised, the message as a whole is protected.

Physical distance between the discovered paths in a multipath routing protocol is critical in the case of wireless networks. This is because of the mutual interference between the paths that can degrade the communication quality. \textit{Spatially disjoint multipath routing protocol} (SDMR) is proposed in \cite{galvez2008spatially}. A source node floods route request packets towards its destination. On the way back, in the Route Reply packets, the intermediate nodes, between the source and the destination, embed their link-state information (containing information about one-hop neighbors) in these packets. This helps the source node to obtain a partial graph of the whole network and choose paths that have the largest \textit{spatial disjointness}. The packets are then forwarded using source routing in a round robin manner over the discovered multiple spatially disjoint paths. The goal of this scheme is to reduce mutual interference between the discovered paths to gain the benefits of load balancing and bandwidth aggregation in MANETs.

\vspace{2mm}
\textit{Source-Routed Multipath Protocols for WMNs:}
\vspace{1mm}

In \cite{tsai2007interference}, reliability and interference minimization are studied in a multiradio multichannel WMN. In this work, non-disjoint multiple paths are considered, and reliability is achieved through packet duplication over these paths. A metric, called \textit{weighted interference minimization path selection metric (WIM)}, is introduced for interference minimization among the paths. This metric and the accompanying algorithm consider interference among paths and neighboring nodes using the same radio channel. It has been shown that this technique, where non-disjoint multiple paths are used, enhances reliability, interference minimization and end-to-end delay as compared to the techniques where only maximally disjoint paths are considered.

In \cite{tam2007joint}, \textit{joint multi-channel multi-path control} (JMM), which is a cross-layer design involving the multi-channel link layer and the routing network layer, is proposed to enable multipath routing in WMNs. In this study, the time horizon is divided into slots, and then different channels are attributed to different time slots. By this way, the contending input traffic is balanced at different times over different channels and different paths. As a result, the end-to-end throughput performance is enhanced.

\subsection{Hop-by-Hop Routing}
\label{subsec:hop-by-hop}

In a hop-by-hop routing scenario, routing decisions are taken, as opposed to source routing, in a distributed or per-hop fashion. Each node along a path processes incoming packets and forwards them by consulting a routing table stored in its memory. Hop-by-hop routing is especially popular for multihop and infrastructure-less environment such as WSNs due to its decentralized and robust nature. Hop-by-hop is by far the most famous and widely adopted technique in IP networks. Link state (LS) and distance vector (DV) routing schemes are the most popular techniques that have been adopted in such routing scenarios. In the rest of this section, we review works that deploy this routing technique in different network types.

\vspace{2mm}
\subsubsection{Hop-by-Hop Multipathing (General/ Wired Solutions)}
\label{subsub: w_g_hop}

A \textit{link state}-based algorithm is proposed by Vutukury and Aceves in \cite{vutukury1999simple} to calculate loop-free multiple paths with minimum dependance on the underlying physical network topology. In \cite{vutukury1999algorithm} and \cite{vutukury2001mdva}, Vutukury and Aceves present a \textit{distance vector}-based algorithm to address the same problem of finding loop-free multiple paths. In all of these works, the multiple paths are provisioned based on a special condition called \textit{loop-free invariant} (LFI) (which is described in \cite{vutukury1999simple}). When a routing algorithm satisfies this condition, then the loop-freeness at every instance is ensured.

In \cite{ohara2009mara}, a family of protocols called \textit{maximum alternative routing algorithm} (MARA), which is based on a directed acyclic graph (DAG), is presented. Using the DAG approach ensures that loop-free paths are constructed. Based on a network topology, the DAG approach constructs a graph with the maximum number of multiple paths within a minimum amount of time subject to a throughput constraint in order to improve the reliability of communication.

\textit{Hashing} enables a node to select a path from a set of available paths while forwarding packets along a path. Two hashing schemes, namely \textit{direct hashing} and \textit{table-based hashing}, are studied in \cite{cao2000performance} for the purpose of packet forwarding in a multipath routing scenario. In direct hashing, a hash function is performed on a packet's header, and the output is directly mapped to one of the paths from the set of constructed multiple paths. In table-based hashing, on the other hand, the output of a hash function is first put in one of the bins (with the total number of bins potentially differing from the total number of available paths) and the bins are then mapped to multiple paths. In the work of Cao et al. \cite{cao2000performance}, five 16-bit cyclic redundancy check (CRC)-based direct hashing techniques and an XOR of source-destination IP-based and table-based hashing technique are developed for balancing the load over \textit{equal-cost multiple paths} (ECMP). ECMP is a simple method for provisioning multiple paths that has been incorporated into many conventional routing environments (such as OSPF, MPLS and ISIS) \cite{villamizar1999ospf, villamizar1999mpls}. ECMP is explained in detail in Section \ref{subsubsec:ecmp}. An interesting feature of table-based hashing is that it can be used to balance load over unequal path weights (as described in \cite{cao2000performance}) at the cost of requiring more states than direct hashing.

\textit{Dispersity routing} enables concurrent transmission of data over multiple paths \cite{maxemchuk1975}. There are two different flavors of dispersity routing. A node can either send distinct data packets over multiple available paths for load balancing, or send duplicated data packets over multiple paths for improved reliability and resiliency.

R-BGP is an extended version of the traditional BGP routing in which multipath routing is applied to hierarchical networks comprising different ASes \cite{kushman2007r}. In addition to the primary paths, the additional multiple paths serve as backup paths, and all of them aim to maintain connectivity even in the face of multiple failures subject to the compliance with AS policy. In  \cite{wang2009path}, an advertisement mechanism is proposed for BGP to advertise information regarding multiple paths at the AS level \cite{wang2009path} while minimizing control overhead with the objective of providing path diversity as a means to circumvent failures. In \cite{van2009loop}, BGP is enhanced to calculate loop-free multiple paths.

\textit{Yet another multipath routing} (YAMR) is a multipath routing protocol that constructs multiple paths and establishes concurrent transmission in hierarchical networks \cite{ganichev2010yamr}. It renders reliable communication against any single interdomain link failure. Compared to BGP, it incurs lower control overhead by isolating failures and keeping the whole Internet oblivious of their occurrences. This technique of containing the failure is termed as \textit{hiding}. This technique combined with multipathing helps YAMR to converge with a lower control overhead than BGP while demonstrating enhanced reliability.

\vspace{2mm}
\subsubsection{Hop-by-Hop Multipathing (Wireless Solutions)}

In this section, we will present a summary of hop-by-hop multipathing protocols that have been proposed for wireless networks.

\vspace{2mm}
\textit{Hop-by-Hop Multipath Protocols for WSNs:}
\vspace{1mm}

Conventional single-path routing schemes put all traffic on a precomputed set of links, thus quickly draining the energy of the intervening network elements. In multipath routing techniques \cite{ming2007energy}, on the other hand, network traffic is distributed over multiple paths with load balancing and concurrent transmission, thereby avoiding overburdening a single set of network elements and enhancing the network lifetime (a critically important feature in WSNs).

The provisioning of multiple paths in a wireless scenario is prone to intra- and inter-path interference. In \cite{maimour2008maximally}, a radio disjoint multipathing method is proposed. This technique constructs multiple paths in an incremental fashion keeping in view the congestion and interference effects, so it helps to improve bandwidth gains among bandwidth-hungry multimedia applications. In the first instance, only one path is established; and upon receiving feedback on network congestion, other paths are established in an incremental fashion. However, when high interference is observed, sensor nodes may be put into passive mode to reduce the number of multiple paths, and with that the amount of interference as well. Overall, this technique improves throughput and energy efficiency but at the cost of higher control overhead.

Another technique called \textit{meshed multipath routing} (M-MPR) with \textit{selective forwarding} is proposed in \cite{de2003meshed, de2003WCNC} to incorporate forward error correction (FEC) and selective forwarding into multipath routing for better reliability in order to improve throughput performance. In selective forwarding, a packet is forwarded over the path with the best downstream wireless channel quality, with acknowledgement driven retransmissions eliminated by using FEC mechanism, which facilitates efficient load balancing with lesser control overhead. It has been shown that M-MPR provides higher throughput than its \textit{disjoint multipath} counterparts.

Yang et al. \cite{yang2010network} have proposed a \textit{network coding-based reliable disjoint and braided multipath routing} protocol. In the disjoint routing approach, multiple disjoint paths are created in a hop-by-hop fashion and routing information is stored locally on a node which does not take end-to-end path in view. A node groups its neighbors based on their respective distances from a destination node in terms of the number of hops hence aiding the load balancing process, in which each packet is forwarded after an election at a node with the path with lesser number of hops towards a destination being selected. Each node calculates a few logical paths as backups, which may not be completely disjoint, and hence are called braided multiple paths. Reliability is introduced in this work by the deployment of network coding at the source and the intermediate nodes. The encoded packets using this technique can only be decoded at the respective destination nodes. Simulation results show that multipathing with braided paths works better in terms of reliability and load balancing in WSNs. Braided multiple paths are considered in another study \cite{ganesan2001highly} for improving reliability and energy efficiency. It is observed that the braided path technique can avoid 50\% more isolated failures while also having a lesser control overhead when compared to disjoint path techniques.

A geographical approach known as \textit{geographical power efficient routing} (GPER) for multipathing has been proposed for WSNs \cite{wu2004gper}. This protocol enables individual nodes to select their own transmission ranges depending upon their respective residual energy resources thus enhancing network lifetime. This act also helps in preserving the energy resources of a passive listening neighbor. As a further improvement, multiple paths are provisioned on a per-node basis. This approach is adopted so that the overall communication can be distributed over a wider area of a network and hence saving a particular set of network elements from being overwhelmed and deprived of their energy resources sooner than the rest of the network. It has been observed that by doing so enhances the network lifetime by up to 30\%.

Although the traditional focus of WSN research has been on energy efficiency, the requirement of security is also very important---especially for environments such as military or healthcare. Stavrou et al. \cite{stavrou2010survey} have surveyed security-related issues. A threat model that covers the behaviors of malicious users is presented, and the secure multipath routing protocols for WSNs are categorized according to their specific operational capabilities in achieving security-related objectives.

The problem of joint fault and intrusion tolerance is considered in \cite{challal2011secure}. A distributed and in-network fault-intrusion tolerance protocol called \textit{secure and efficient intrusion-fault tolerant protocol} (SEIF) is developed. This technique renders nodes to take decisions in regards to resilience and reliability that come with independence from base station. By this way, the time incurred to detect an intrusion is reduced. This security mechanism is incorporated into a multipath routing protocol called \textit{sub-branch multipath routing protocol} (SMBR) to improve network discovery time and energy efficiency, as well as to reduce control overhead as only one message per node is required to construct a network topology. H-SPREAD \cite{lou2006h} is a routing protocol for WSNs that facilitates the construction of multiple node-disjoint paths towards the base station in a distributed manner. In such a setting, only a few paths can be compromised if some malicious nodes appear in the network.

\vspace{2mm}
\textit{Hop-by-Hop Multipath Protocols for MANETs:}
\vspace{1mm}

\textit{Ad-hoc on-demand multipath distance vector routing} (AOMDV) \cite{marina2001demand} is one of the famous modifications of AODV. AOMDV provisions multiple link-disjoint paths between a source-destination pair in an on-demand manner. Another AODV modification is presented  by Ye et al. in \cite{ye2003framework}. Reliable nodes, in terms of energy resources, are deployed in the network topology. These nodes are used to construct reliable multiple paths, along with other unreliable nodes, to ensure routes are more reliable and less prone to failures and outages in MANETs. An algorithm is presented to control the placement and trajectories of the reliable nodes so that these nodes are highly likely to be part of a path that a normal (or unreliable) node constructs. An important discovery made in this work is that there can only be a few disjoint paths between a source-destination pair in a MANET---with the number of disjoint paths decreasing as the distance between communicating pairs increases. This discovery motivates the use of reliable nodes in such networks.

 \textit{MP-AODV} is an adaptive multipath routing scheme based on the conventional AODV routing scheme proposed in \cite{sambasivam2004dynamically}. Primary paths are selected based on the received signal strength. Alternative paths are then constructed by the exchange of, so called, source generated \textit{heartbeat messages} among nodes in a MANET. These messages help a source to gather the mobility information of nodes that make up the multiple (backup) paths towards a destination. One of the main objectives of this work is to enhance the reliability of communication in MANETs. Another work \cite{li2004demand} presents \textit{node-disjoint multipath routing} (NDMR) protocol that also makes use of backup paths, and improves upon the legacy protocols, namely \textit{AODV} and \textit{DSR}. In this work, the issue of reliability in terms of \textit{packet delivery ratio} is addressed. Besides reliability, this protocol is also efficient in terms of end-to-end delay and control overhead.

A technique called \textit{packet-caching} is proposed by Valera et al. in \cite{valera2003cooperative} to ameliorate congestion and packet losses with the provision of multiple paths. In this work, \textit{caching and multipath (CHAMP)}, which consists of cooperative packet caching and shortest multipath routing, is proposed to cater for route breakdown, which is a common yet important issue in MANETs. In CHAMP, each node has a buffer in which it stores/ caches data packets that it encounters during a normal operation. Information of alternate routes is also cached at the same node to be used when the downstream node (relative to this node) fails to forward a data packet or somehow malfunctions. If such a situation occurs, an upstream node forwards the dropped packet from its cache along an alternative path. This technique performs better than DSR and AODV in terms of end-to-end delay, packet delivery ratio and control overhead. In another work \cite{zhang2010reliable}, \textit{reliable adaptive multipath provisioning (RAMP)} is proposed for reliable adaptive multipath provisioning to meet bandwidth and differential delay constraints. The problem of differential delay is particularly important for the multipath routing scenario. When traffic is dispersed over multiple paths, then different paths may offer different amount of delays, and as a result the traffic reaches at a destination at different times. This difference in delay among multiple paths is called the differential delay. Traffic is allowed to traverse over multiple paths, which may be disjoint, so that communication outage due to a single link failure can be avoided.

An energy saving protocol for MANETs is proposed by Srinivas and Modiano in \cite{srinivas2003minimum}. In order to conserve energy, 2-link-disjoint and k-node-disjoint algorithms are proposed for dense and sparse networks, respectively. A set of algorithms are developed to find node- or link-disjoint paths in a given network graph so that transmission energies of nodes are minimized. An interesting observation has been reported in this work, that is the incremental energy of each newly added link-disjoint path is decreasing due to the wireless nature of the underlying network. 

There has also been work done on multipathing for a specific configuration of MANET known as vehicular ad-hoc networks (VANETs). In \cite{huang2009performance}, a node-disjoint concurrent multipathing protocol is proposed for VANETs to improve the overall packet delivery ratio even though the relaying nodes may have unknown speeds. In another video streaming work for VANETs \cite{wang2012liaithon}, \textit{location-aware multipath video streaming} (LIAITHON) is proposed for urban vehicular networks. In this protocol, a receiver initiates a video streaming process by sending a request to a source. The source then transmits the video stream over the intermediate nodes that are well away from each other to reduce inter-path interference. The selection of such nodes is carried out using the geographical location information of these nodes (with respect to the source node).

There have also been biologically inspired routing protocols proposed in MANETs, particularly AntHocNet, which is an \textit{ant colony optimization} (ACO)-based multipath routing protocol \cite{di2005anthocnet, ducatelle2005ant}. The ACO-based AntHocNet protocol has both a proactive path probing component as well as a reactive path setup/maintenance component. It is shown in \cite{di2005anthocnet, ducatelle2005ant} that AntHocNet outperforms AODV in terms of delay, jitter and packet delivery ratio, while having the drawback of being more expensive in terms of its control overhead.

\vspace{2mm}
\textit{Hop-by-Hop Multipath Protocols for CRNs:}
\vspace{1mm}

In \cite{beltagy2011new}, a multipath routing scheme for cognitive radio networks (CRNs) is deployed by introducing a metric that tells about the \textit{route closeness} of multiple paths, which depends on the physical distance among the multiple available paths. In CRNs, an unlicensed or secondary user (SU) can access the licensed radio spectrum if it is not being used by the licensed or primary users (PUs). Since the PU has higher priority over SUs, the arrival of PU precludes SUs from accessing the spectrum and makes the routing problem challenging. The paths with less closeness are preferred because such paths are less prone to the PU activities due to the greater distance among these paths. In another interesting CRN-based work \cite{ali2015}, contrary to most of the multipath routing works presented in this paper (that have addressed unicast routing), Ali et al. have proposed a \textit{multipath-based approach to multicast routing} in CRNs. The proposed approach makes use of learning automata to select active and backup paths to develop multicast routing protocols that is minimally disrupted by PU activity. 

\vspace{2mm}
\textit{Hop-by-Hop Multipath Protocols for PANs:}
\vspace{1mm}

There have also been works done on multipathing for wireless personal area networks (WPANs). Multipathing is realized over the MAC layer of the low-rate WPAN (LR-WPAN) standard IEEE 802.15.4 in \cite{pavkovic2011multipath}. This realization is achieved by selecting multiple parent nodes in an opportunistic manner at a time. This is done by making IEEE 802.15.4's MAC layer's cluster tree to operate in harmony with the \emph{destination oriented directed acyclic graph} (DODAG) as proposed in \emph{routing protocol for low-power and lossy networks} (RPL) \cite{hui2012routing}.

\subsection{Overlay Routing}
\label{subsec:overlay_routing}

We have earlier seen that source routing equips end hosts with the ability to deal with any failures arising along a route itself. The source can thus either identify an end-to-end route itself as is done in the ``source routing'' paradigm, or it may use the ``overlay routing'' paradigm to choose some intermediate hosts to route through in order to avoid performance and reliability problems that may exist on the direct route. An overlay network is essentially a virtualized logical network built on top of a physical network with tunnels interconnecting edge devices. The overlay network appears to the nodes connecting to it as a native network with the possible existence of multiple overlays on the same underlying physical infrastructure (which allows support for multi-tenancy). Overlay routing can be used to override suboptimal routing in the substrate networks, and can be used to support more flexible self-driven routing control of multihomed end hosts. In this section, we will discuss the overlay routing paradigm.  


\vspace{2mm}
\subsubsection{Overlay Routed Multipathing (General/ Wired Solutions)}
\label{subsub:w_g_overlay}

Source routing is not supported in most of the Internet routers. The most common technique used to support multipath on the Internet is to maintain an overlay network for connecting nodes and supporting source routing. Overlay routing is especially popular on the Internet for supporting TE and multipathing.

Multihoming for overlay networking is deployed in \cite{andersen2005improving}, and a new \textit{multihomed overlay network} (MONET) protocol is proposed. The purpose of this work is to increase the availability of websites to users by creating multiple paths to web proxies. A method called \textit{way point selection} is used to choose a subset of available paths leading to the proxies in order to access websites. This method decides the order in which each path is chosen to forward packets towards a particular destination server depending upon the success statistics of the paths. In another work, Andersen et al. \cite{andersen2001case} proposed to enable an end user to switch between actual network paths and those constructed by an overlay network. This incentive is also available to the applications depending on their network requirements. In order to enhance bandwidth availability to meet application requirements, an overlay networking-based multipath solution is presented in \cite{lee2008bandwidth}.

Multimedia streaming is another important application in which reliability is important for seamless transmission. Multipathing reduces jitter and packet loss for efficient and reliable multimedia streaming. A coding technique called \textit{multiple description coding} (MDC) \cite{apostolopoulos2000reliable}, \cite{apostolopoulos2002multiple} is deployed to enable multipathing for streaming media. In this coding technique,  streaming media from content distribution networks is divided into multiple independently decodable streaming flows \cite{goyal2001multiple}. Packets from these multiple streaming flows are then transmitted over multiple disjoint paths. Many works base their multipath techniques on the streaming media using MDC.

In another overlay routing scheme, Savage et al. proposed the \textit{Detour} overlay network \cite{savage1999end}---comprising geographically distributed routes interconnected using tunnels (which  act as virtual point-to-point links) over the commodity Internet---to be used for better end-to-end performance by routing around performance bottlenecks. Any host using the Detour network directs its outbound traffic to the nearest Detour router with the packets taking an exit at a point close to the destination. Detour uses a dynamic multipath routing approach with automatic load balancing and congestion avoidance for high performance. The router and congestion is controlled at the Detour routers---which are, technically speaking, still edge devices of the overlay network.

In \cite{ma2004new}, a quality of service (QoS) metric based on \textit{link correlation} is presented for overlay networks. This metric, along with \textit{path correlation}, is then used to select multiple paths for an MDC streaming video flow. It is observed that this overlay network-based technique improves SNR compared to the conventional link-disjoint methods by up to 3.2 dB. Another work in \cite{begen2005multi} makes use of overlay routing and MDC for efficient and reliable video streaming. Based on MDC, different multipath streaming models have been derived. Then, based on these models, an optimum strategy is given to select an optimum set of multiple paths. This technique also improves the SNR by up to 6.07 dB as compared to the single-path and link-disjoint multipathing.

Various researchers have studied how to support virtual overlay networks on top of the physical substrate networks, with a few works exploiting multipath for efficiently supporting virtual networks (VN) on the underlying substrate. Yu et al. \cite{yu2008rethinking} have studied the strategies that the substrate network can adopt to flexibly support virtual network embedding in which the virtual network requests are mapped to specific nodes/ links in the underlying substrate network. Yu et al. simplify the general problem of virtual link embedding (which is known to be computationally intractable) by allowing the substrate network to (1) split a virtual link over multiple substrate links, and (2) migrate paths to periodically optimize the utilization of substrate network. The related problem of multi-domain VN embedding with limited information disclosure is studied by Dietrich et al. \cite{dietrich2013multi} who proposed a VN embedding framework that enables VN request partitioning scheme while assuming only limited information disclosure.

\vspace{2mm}
\subsubsection{Overlay Routed Multipathing (Wireless Solutions)}

In this section, we introduce various overlay multipath routing protocols proposed for wireless networks, specifically WSNs.

\vspace{2mm}
\textit{Overlay Multipath Protocols for WSNs}
\vspace{1mm}

A multipath routing protocol called MMSPEED \cite{felemban2006mmspeed} is proposed to achieve QoS guarantee. In this protocol, multipath routing is deployed in a distributed overlay manner to improve reliability while satisfying the constraint of packet delivery speeds. Here, the packet delivery speed is equal to the ratio of one-hop average distance to one-hop average delay. Braided multiple paths are considered in \cite{ganesan2001highly} for the enhancement of reliability and energy efficiency. It has been observed that this technique works 50\% better as compared to its \textit{disjoint} counterparts in terms of isolated and patterned failure avoidance with low control overhead. The isolated failure here refers to those incidents in which nodes fail randomly and independently, while in the case of patterned failures, nodes belonging to a specific geographic region fail.

\begin{table*}[!ht]
\centering
\scriptsize
\ssmall
\caption{Routing Metric Based Classification of Multipath Routing Frameworks}
\label{tab:metrics}
\begin{tabular}{p{11.8cm}p{5.9cm}}
\hline

\cellcolor[HTML]{EFEFEF}\textbf{\textit{Routing Metric}}
& \cellcolor[HTML]{EFEFEF}\textbf{\textit{Sample Works in Various Networking Configurations}}\\
\hline
\\

\multicolumn{2}{l}{\cellcolor[HTML]{EFEFEF} \textbf{\textit{Multipath Specific Routing Metrics}}}\\
\\

\textbf{\textit{Path Disjointness/ Independence}}: To gain the benefits of multipath routing, it is important to choose paths that are, as much is possible, independent from each other. With disjoint/ diverse paths, routing is more resilient to failure, and the paths will not share a common bottleneck. & \textit{General/ Wired}: \cite{sidhu1991finding}, \cite{guo2003link} ; \textit{WSNs}: totally disjoint paths \cite{waharte2006totally}, maximally radio disjoint routing \cite{maimour2008maximally}; \textit{MANETs}: SDMR \cite{galvez2008spatially}, SMR \cite{lee2001split}; \textit{VANETs}: Node-disjoint-multipath \cite{huang2009performance}, LIAITHON \cite{wang2012liaithon}.
\\

\\
\textit{\textbf{Path Closeness/ Correlation}}: Similar to the schemes that aim at building link/ node-disjoint paths, routing protocols can choose the metric of `path closeness' or `path correlation' to prefer diverse paths over paths that are correlated. Paths that are more dissimiliar and independent tend to provide better performance. & \textit{General/Wired}: \cite{ma2004new}; \cite{liao2011introducing}; \textit{MANETs}: SDMR \cite{galvez2008spatially};\newline \textit{VANETs}: LIAITHON \cite{wang2012liaithon}; \textit{CRNs}: \cite{beltagy2011new}.\\

\\
\textbf{\textit{Interpath Interference Minimization}}: Routing protocols can be designed to prefer paths that minimize interpath interference. With reduced interpath interference, concurrent usage of resources can be performed and the overall routing performance improves. & \textit{General/ Wired}: Dynamic Constrained Multipath Routing with MPLS \cite{seok2001dynamic} ; \textit{WMNs}: weighted interference minimization (WIM) metric \cite{tsai2007interference}; \textit{WSNs}: \cite{maimour2008maximally}, IM2PR \cite{radi2014im2pr}.\\

\\
\multicolumn{2}{l}{\cellcolor[HTML]{EFEFEF} \textbf{\textit{Traditional Routing Metrics Used For Multipath}}}\\
\\

\textbf{\textit{Delay/ Latency}}: Delay refers to the average end-to-end latency experienced by a communicating source-destination pair. We are interested in minimizing latency which can be achieved by minimizing network congestion and alleviating bottlenecks. Solutions include: concurrent multipathing for dispersing the traffic to avoid bottlenecks, and using backup paths for more rapid route rediscovery after failure. &  \textit{General/Wired}: \cite{vasic2010energy, vutukury2001mdva};\textit{WSNs}:  \cite{felemban2006mmspeed, ming2007energy, huang2008multiconstrained, yahya2009reer}; \textit{WMNs}: \cite{tsai2007interference}; \textit{MANETs}: \cite{lee2001split, tsirigos2001multipath, li2004case, li2004demand, chen2005multipath, di2005anthocnet, ducatelle2005ant, zhang2010reliable, marina2001demand}; \textit{Other}:  \cite{huang2009performance, pavkovic2011multipath, gui2012distributed}.\\

\\
\textbf{\textit{Reliability/ Resiliency}}: Reliability refers to the ability of the networking protocols to work consistently as required. Failures can be circumvented by using backup paths, and higher packet delivery ratio can be ensured through concurrent multipathing over diverse links. & \textit{General/Wired}: \cite{gummadi2004improving, andersen2005improving,  xu2006miro, kushman2007r, yang2007nira, motiwala2008path, wang2009path, godfrey2009pathlet, ganichev2010yamr}; \textit{WSNs}: \cite{felemban2006mmspeed, lou2006h, huang2008multiconstrained, yahya2009reer, yang2010network, challal2011secure, wang2000multipath, karlof2003secure}; \textit{WMNs}:  \cite{tsai2007interference}; \textit{MANETs}: \cite{tsirigos2001multipath, ye2003framework, valera2003cooperative, lou2004spread, li2004case, li2004demand, chen2005multipath, di2005anthocnet, ducatelle2005ant}; \textit{CRNs}: \cite{beltagy2011new}; \textit{VANETs}: \cite{huang2009performance}.\\

\textit{\textbf{Bandwidth/ Throughput}}: Throughput, or more informally bandwidth, refers to the amount of data that can be transferred within a period of time. The use of traditional single-path routing schemes can result in congestion hotspots that act as a performance bottleneck. Load balancing and dispersion of network traffic using multiple paths at the same time offer higher bandwidth and throughput gains. & \textit{General/Wired}: \cite{van2009loop}; \textit{WSNs}: \cite{de2003meshed, de2003WCNC, waharte2006totally, nasser2007seem, ming2007energy, maimour2008maximally, wang2000multipath}; \textit{WMNs}: \cite{tam2007joint, nandiraju2006multipath, radunovic2008optimization}; \textit{MANETs}: \cite{li2004case}; \textit{CRNs}: \cite{wang2009multipath, khanna2011interference}.\\
\\
\textbf{\textit{Control Overhead}}: Control overhead refers to the number of control messages that need to be exchanged to compute and maintain multiple paths. An efficient multipathing protocol minimizes control overhead. & \textit{General/Wired} \cite{vutukury2001mdva}; \textit{WSNs}: \cite{nasser2007seem, wang2000multipath}; \textit{WMNs}: \cite{lee2000aodv, radunovic2008optimization}; \textit{MANETs}: \cite{lee2001split, marina2001demand, wu2001performance, nasipuri1999demand}; \textit{CRNs}: \cite{khanna2011interference}.\\

\\
\textbf{\textit{Energy Efficiency}}: Energy efficiency refers to the extent the multipath technique conserves energy resource. Such a metric is important for WSNs and MANETs in which energy is a limited and scarce resource. Both concurrent transmission and traffic engineering can increase network lifetime by dispersing the traffic over multiple paths.  & \textit{General/Wired}: \cite{valiant1981universal, valiant1982scheme, Valiant1990, zhang2005designing, zhang2008designing, zhang2010valiant} ; \textit{WSNs}: \cite{dulman2003energy, liu2007pwave, ganesan2001highly, de2003meshed, de2003WCNC, chen2006energy, wu2007power, nasser2007seem, ming2007energy, maimour2008maximally, yahya2009reer, yang2010network, challal2011secure}; \textit{WMNs}: \cite{akyildiz2005wireless}; \textit{MANETs}:  \cite{srinivas2003minimum, kun2005research, wu2001performance}; \textit{CRNs} \cite{gui2012distributed}.\\


\\
\hline
\end{tabular}
\end{table*}

\subsection{SDN-based Routing}
\label{subsec:sdn_based}

In this section, we discuss SDN-based multipath routing proposals. SDNs allow networks to be programmable through the separation of the control plane and the data plane, as well as through the support of high-level control abstractions \cite{qadir2014prog}, \cite{qadir2014acmfit}. The centralized decision making approach at the SDN controller, along with the development of new sophisticated control abstractions, allows convenient control and management of networks, and can be used to implement multipath routing. 

\subsubsection{SDN-Routed Multipathing (General/ Wired Solutions)}
\label{subsub: w_g_sdn}

Software-defined networking (SDN) is becoming increasingly popular for management of networks, especially in DC environments. SDN is often an enabling technique for the operations of DCs. Rain Man is a SDN-based firmware architecture proposed in \cite{stephens2012designing}. This firmware has a layer two (L2) routing mechanism that enables efficient multipath routing in DC networking. Another work called Hedera, with a global view of active flows, presents an adaptive and efficient multipathing approach based on SDN concepts \cite{al2010hedera}. In \cite{benson2011microte}, Micro-TE is proposed so that flows in a DC environment are scheduled using a central controller with the help of stochastic properties of network traffic and multipath routing. Full link utilization is achieved in B4 \cite{jain2013b4} using traffic engineering solution with the help of OpenFlow-enabled centralized controller and packet forwarding mechanism using multiple paths. OpenFlow is a protocol for SDNs through which the control plane of a central controller talks to (or programs) the data plane of the switches in the network.

\vspace{2mm}
\subsubsection{SDN-Routed Multipathing (Wireless Solutions)}

In this section, we introduce various overlay multipath routing protocols proposed for the wireless networking configurations of MANETs, WMNs and VANETs.

\vspace{2mm}
\textit{SDN-based Multipath Protocols for MANETs}
\vspace{1mm}

SDN concepts have been used in \cite{kusoftware} to realize a MANET infrastructure. It has been shown that carrying out routing operation through SDN enhances the reliability of such networks. Concurrent transmission of data packets has been studied as a use case to demonstrate that this technique increases packet delivery ratio at the expense of moderate control overhead increase.

\vspace{2mm}
\textit{SDN-based Multipath Protocols for WMNs}
\vspace{1mm}

An interesting study called \textit{wireless mesh software-defined networks (wmSDN)} is presented in \cite{detti2013wireless}. In this study, a mesh network comprises wireless mesh routers (WMRs) that are IEEE 802.11-based and OpenFlow-enabled. The data plane of these WMRs is programmed via a centralized controller. Traffic balancing is carried out by the central controller using an algorithm called \textit{gateway selection algorithm (GSA)}. In this technique, the WMRs act as gateways for mesh network to forward traffic out of the network, and these are selected for Internet connectivity in a round robin fashion, where the least used gateway is selected first and so on, instead of using the nearest gateway at all times to forward traffic out of this network. This is done in order to achieve an enhanced network utilization. 

\vspace{2mm}
\textit{SDN-based Multipath Protocols for VANETs}
\vspace{1mm}

A similar work is presented for the case of VANETs in \cite{kutowards}. In this work, routing decisions are controlled using a central controller in an SDN environment. Multipathing provides resiliency in the case of emergency scenarios, such as route failures, or for the high priority flows. To achieve this, the controller reserves a few paths dynamically as backups.

\subsection{Routing Metrics for Multipath Routing}
\label{sec:metric_cat}

Different metrics have been used to quantify the performance and quality of multipath routes, including some that are direct extensions of the common routing metrics used in single-path routing (e.g., throughput, delay, congestion) and others that are customized for the unique settings and requirements of multipath routing. We discuss both types next. A summary of the various metrics used for multipath routing is presented in Table \ref{tab:metrics}. The choice of routing metrics used by various wired and wireless multpathing works is also summarized in Tables \ref{tab:MultipathRouting} and \ref{tab:MultipathRoutingWireless}. 



\vspace{2mm}
\subsubsection{Multipath Specific Metrics}

The quality of multipath routing essentially depends on the independence and diversity of the paths chosen (so that an isolated fault/ failure is not able to affect all of the paths). We have divided multipath specific metrics into three types, namely those that measure: (i) path disjointedness/ independence; (ii) interpath interference; and (iii) path closeness/ correlation. These metrics are related to each other---e.g., if we select multiple paths that are disjoint, we also select paths with lesser closeness/correlation. The similarity of these metrics stems from the fact that they all measure correspondence to the qualitative ideal of maximally independent/diverse collection of paths that do not share a common bottleneck. Each of these metrics has been used in various network configurations including wired and wireless networks (as can be seen in Table \ref{tab:metrics}). 

\vspace{2mm}
\textit{Path Disjointedness/ Independence}: As noted earlier, the key to benefiting from a multipath routing framework is to ensure that multiple paths support each other when needed. This is best ensured when all the paths are independent so that a congestion hotspot or a performance bottleneck is not able to simultaneously affect the different paths. Many proposed multipath routing works have utilized some variation of this metric for path computation to measure link- or node-disjoint paths that are maximally independent. Examples of works proposed for wired networks include \cite{sidhu1991finding}, \cite{guo2003link}; while works proposed for wireless networks include totally disjoint paths \cite{waharte2006totally} and maximally radio disjoint routing \cite{maimour2008maximally} for WSNs; SDMR \cite{galvez2008spatially} and SMR \cite{lee2001split} for MANETs, and node-disjoint multipath and LIAITHON \cite{wang2012liaithon} for VANETs.

\vspace{2mm}
\textit{Interpath Interference Minimization}: Routing protocols can be designed to prefer paths that minimize interpath interference. With reduced interpath interference, concurrent usage of resources can be performed and the overall routing performance improves. A number of interpath intereference minimizing multipath protocols have been proposed, both for wired networks (such as the MPLS-based dynamic constrained multipath routing approach \cite{seok2001dynamic}) as well as for wireless networks (such as the use of weighted interference minimization or WIM metric for WMNs \cite{tsai2007interference} and the IM2PR protocol for WSNs \cite{radi2014im2pr}).

\vspace{2mm}
\textit{Path Closeness/ Correlation}: Path closeness/ correlation is very similar to path disjointedness/ independence metric---and indeed, it can be thought of as its variant. With multipath routing, we seek to minimize the metrics of `path closeness' or `path correlation' to prefer multipath paths that are mutually more diverse. Such metrics have been used both for wired networks \cite{ma2004new}, \cite{liao2011introducing} as well as for wireless networks (such as the SDMR \cite{galvez2008spatially} work for MANETs, and the work of Beltagy et al. for CRNs). 

\vspace{2mm}
\subsubsection{Traditional Routing Metrics}

Some applications might be delay sensitive, while the others put emphasis on reliability. Here, we present different classes of multipath routing protocols whose design considerations revolve around a specific metric. The metrics being discussed in this section include: (i) delay, (ii) throughput, (iii) reliability, (iv) control overhead, and (v) energy efficiency. 

\vspace{2mm}
\paragraph{Delay/ Latency}
\label{subsec:delay}

In many time-sensitive applications, the delay metric is of vital importance. In the case of WSNs, time sensitivity matters in the case of military, health and life-saving scenarios, such as a fire-detection network application. In such scenarios, using multiple paths concurrently can significantly reduce end-to-end delay. In concurrent multiple-path transmissions, which serve as congestion avoidance mechanism \cite{villamizar1999ospf}, \cite{villamizar1999mpls}, using efficient load balancing and flow-splitting techniques help to avoid bottlenecks contributing to lower end-to-end delay. Another aspect is in the usage of multiple paths as backups \cite{li2004demand}, \cite{sambasivam2004dynamically}, \cite{marina2001demand}. In this case, when a primary path fails, the traffic is shifted to one of the alternative paths from a pre-calculated set of backup paths. In this way, the amount of time for route rediscovery is reduced.

\vspace{2mm}
\paragraph{Bandwidth/ Throughput}
\label{subsec:BW}

Bandwidth efficiency, or efficient utilization of network resources, is among the most important benefits of multipathing. Concurrent multipathing, with efficient techniques of load balancing and flow splitting, distribute the network traffic across a network. By this way, the network traffic does not overburden a single set of nodes and links as opposed to the case of single-path routing \cite{phatak2002novel}, \cite{maxemchuk1975}, \cite{villamizar1999ospf}. The majority of existing state-of-the-art multipathing protocols utilize bandwidth/ throughput as part of the goals in network performance enhancement. 


\vspace{2mm}
\paragraph{Reliability/ Resiliency}
\label{subsec:reliability}

Packet delivery ratio is an important metric to measure reliability/ resiliency. A number of factors such as network congestion, interference, mobility make links/ nodes prone to failure. Ideally, network routing should be resilient such that in the case of the failure of the primary path, routing switches to an alternative backup path (if available) within a short period of time. In addition, using multiple paths in a concurrent way, the effects of bottlenecks and congestion can be reduced through packet duplication or redundancy. Many works discuss the issue of reliability, resiliency and packet delivery ratio in several networking applications \cite{andersen2005improving}, \cite{xu2006miro}, \cite{kushman2007r}. In the case of general/ wired networks, multipath routing is mostly used to provide reliability in both intradomain and interdomain scenarios \cite{yang2007nira}, \cite{godfrey2009pathlet}. Multipathing provides reliable communications in WSNs \cite{felemban2006mmspeed}, \cite{ming2007energy}, \cite{huang2008multiconstrained}, CRNs \cite{beltagy2011new}, VANETs  \cite{huang2009performance}, IEEE's 802.15.4 \cite{pavkovic2011multipath}, and MANETs \cite{ye2003framework, valera2003cooperative, lou2004spread, li2004demand, chen2005multipath}. 

\begin{figure*}[]
\centering
\includegraphics[width= 0.76\textwidth]{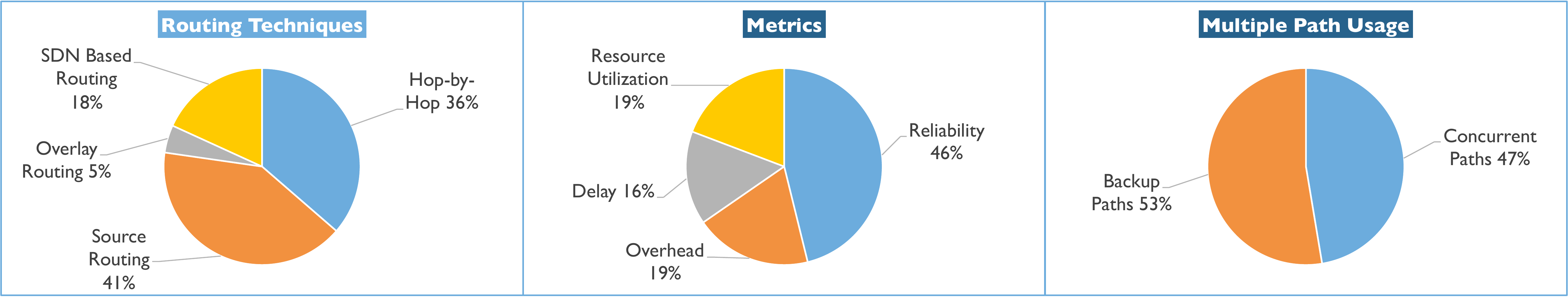}
\caption{Summary of route computation techniques for \underline{\textit{wired}} networks}
\label{fig:infographicWired}
\end{figure*}

\begin{figure*}[]
\centering
\includegraphics[width= 1\textwidth]{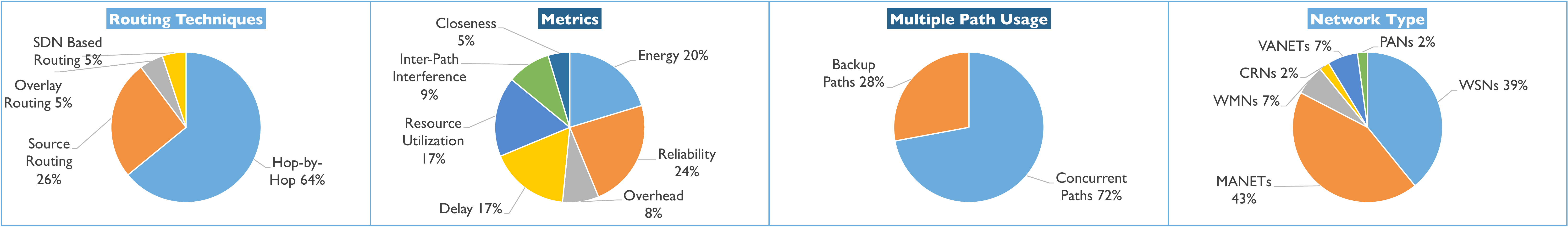}
\caption{Summary of route computation techniques for \underline{\textit{wireless}} networks}
\label{fig:infographicWireless}
\end{figure*}

\vspace{2mm}
\paragraph{Control Overhead}
\label{subsec:controlOverhead}

Implementation of multipath routing is often more expensive than single-path routing due to the extra cost of control messaging imposed by discovering and maintaining more than a single path for a source-destination pair. A number of works have focused on minimizing control overhead associated with multipath routing techniques. An example is a multipath distance vector routing scheme called MDVA shown in \cite{vutukury2001mdva}. In addition to computing loop-free paths, this scheme uses the technique of \textit{diffusing computation}, where a node reports its distance information to an upstream node after calculating it based on the distance information from its downstream node, and neighbor-to-neighbor synchronization which is more efficient in terms of control message exchange compared to distributed Bellman-Ford (DBF) and topology broadcast (TOPB) algorithms; specifically, it improves the convergence time after link failures. The control overhead routing metric is highly relevant to energy efficiency. To avoid drainage of energy resources, it is desirable to contain the number of control messages that need to be exchanged. Researchers have focused on this problem and have proposed routing schemes to limit the control overhead \cite{nasser2007seem, wang2000multipath}. The routing metric of energy efficiency has been applied in WMNs \cite{lee2001split, marina2001demand} and MANETs \cite{lee2001split, marina2001demand}.

\vspace{2mm}
\paragraph{Energy Efficiency}
\label{subsec:Energy Efficiency}

Energy is one of the main concerns in the design of routing protocols, particularly in power-constrained networking configurations such as WSNs and MANETs. Multipath routing, with the use of concurrent multiple paths, improves energy efficiency \cite{ganesan2001highly}, \cite{de2003meshed}, \cite{de2003WCNC}. In a legacy single-path routing, the network resources of the primary path are drained much quicker because of the traffic load it bears. Using multipathing and concurrent transmission with efficient load balancing techniques, the network traffic spreads out ensuring that only a few particular network elements are not overburdened. This provides not only bandwidth and throughput benefits but also energy conservation. 

\subsection{Summary of Trends}
\label{subsec:design_guidelines}

In this section, we identify the main trends observed on the basis of papers presented in this survey (which in our view provide a balanced representative view of the overall multipath routing literature). The trends of the commonly adopted routing approaches/ choices are noted below and summarized visually in Figure \ref{fig:infographicWired} for wired networks, and in Figure \ref{fig:infographicWireless} for wireless networks.   


\begin{itemize}

\vspace{1mm}
\item{Trend 1:} Hop-by-hop routing is the most popularly used route computation technique, whereas overlay routing is the least used (amongst those papers that we have surveyed).


\vspace{1mm}
\item{Trend 2:} Considering the hop-by-hop route computation technique, we note that this multipath routing technique is especially popular in WSNs (compared to other network configurations). This is in contrast to other network types such as wired networks in which hop-by-hop routing and source routing are used a similar number of times. Hop-by-hop routing is, on the other hand, comfortably the most popular choice for WSNs. 



\vspace{1mm}
\item{Trend 3:} Multiple paths can be used in both concurrent/ backup configurations. It is observed (based on the works presented in this study) that wired/ general networks prefer backup and concurrent multiple paths almost equally, while in wireless scenarios, we see that the concurrent transmission option is adopted more often.


\end{itemize}

A multipath algorithm designer can consult Figures \ref{fig:infographicWired} and \ref{fig:infographicWireless} to quickly learn which design choices have traditionally been most popular for the particular networking type or configuration being considered. As an example, if we are to design a multipathing method for WSNs, then the most popular choice is to adopt hop-by-hop routing with concurrent use of the multiple paths while focusing on the energy conservation metric.

\begin{table*}[!ht]
\centering
\scriptsize
\ssmall
\caption{Summary of Flow Splitting/ Load Balancing Techniques}
\label{tab:LoadBalancing}
\begin{tabular}{p{2.7cm}p{1cm}p{13cm}}
\hline

\cellcolor[HTML]{EFEFEF}\textbf{Approach} & \cellcolor[HTML]{EFEFEF}\textbf{Year} & \cellcolor[HTML]{EFEFEF}\textbf{Main Idea} \\

\hline

\\

\multicolumn{2}{l}{\textbf{\textit{Static/ Quasi-Static Approaches}}}\\
\multicolumn{3}{l}{\textit{(In these approaches, the traffic split is \underline{either static} or is managed offline based on a predicted traffic demand matrix with \underline{changes occurring after a singnificantly long period of time)}}}\\

\\
Weighted round robin \textit{or} Deficit round robin \cite{shreedhar1995efficient} & 1995 & Adopts an efficient \textit{fair queuing} method to schedule contending traffic flows in an almost perfectly fair fashion with efficient running time and a complexity of \textit{O(1)} per packet processing time. \\

ECMP \cite{moy1998ospf} & 1998 & Divides traffic evenly over multiple equal-cost paths. It is stable as it does make dynamic adjustments to OSPF costs based on traffic loading. There are three splitting methods in ECMP: 1) Per-flow hashing; 2) Per-packet round robin; and 3) Dividing destination prefixes among the available next hops in forwarding table. \\


Valiant Load Balancing (VLB) \cite{zhang2010valiant}, \cite{Valiant1990} & 1990 & Performs a decentralized 2-phase routing technique, which is agnostic to the traffic matrix. In the first phase, VLB---independent of the destination---redirects traffic to arbitrary intermediate routers. The intermediate routers then forward the traffic to the destination in the second phase.  \\


OSPF-OMP \cite{villamizar1999ospf} & 1999 & Extends OSPF, which can form multiple equal-cost paths and uses ECMP techniques to divide traffics equally over these paths, so that it utilizes opaque LSAs to distribute loading information to facilitate uneven splitting.\\



MPLS Optimized Multipath (MPLS-OMP) \cite{villamizar1999mpls} & 1999 & Proposes a TE procedure to achieve load balancing across multiple label-switched paths in MPLS. In MPLS-OMP, interior gateway protocol (IGP) floods traffic load information according to OSPF-OMP and ISIS-OMP specifications. \\


OSPF/ IS-IS with Link Weight Adjustment \cite{fortz2000internet} & 2000 & Explores near-optimal Internet TE approach under two constraints. Firstly, paths are established using shortest-path routing. Secondly, there is equal flow splitting across equal-cost paths (ECMP) associated with OSPF/ IS-IS for \textit{known demands}. A local search heuristic is proposed for the NP-hard problem of optimizing IGP link weights given a particular traffic matrix. \\


Adapting OSPF/ IS-IS Weights in a Changing Environment \cite{fortz2002optimizing, fortz2002traffic} & 2002 &  Proposes a family of algorithms for efficient OSPF traffic engineering in a dynamic environment. If the traffic demand matrix changes significantly, weight settings are updated with just a few changes.  \\




Sridharan et al. \cite{sridharan2005achieving} & 2005 & Proposes a near-optimal traffic distribution approach based on OSPF/ IS-IS shortest-path routing---which selects a subset of next hops for each prefix using a simple heuristic with provable performance bounds---while preserving the existing protocols and infrastructure.\\

Fong et al. \cite{fong2005better} & 2005 & Proposes techniques for improving OSPF including features that allow flows to traverse non-shortest paths (with a possible cost of occasional loops). \\

DEFT \cite{xu2007deft} & 2007 & Proposes a link-weight-based flow-splitting mechanism that allows routers to redirect traffic on non-shortest paths with an exponential delay on longer paths. DEFT provably improves the performance of OSPF and IS-IS. \\

PEFT \cite{xu2011link} & 2011 & Enhances the link-state routing protocol with hop-by-hop forwarding by utilizing non-shortest paths. Like DEFT, it splits traffic on all paths with longer paths carrying an exponential penalty. PEFT is different from \textit{link-based splitting} methods---such as DEFT \cite{xu2007deft} and Fong et al. \cite{fong2005better}---in that it uses \textit{path-based splitting} (as explained in \cite{xu2011link}).\\

\\
\multicolumn{2}{l}{\textbf{\textit{Dynamic (Load-Aware) Approaches}}}\\
\multicolumn{3}{l}{\textit{(In these approaches, the traffic split is \underline{dynamically} adapted to the changes in observed load)}}\\
\\
MPLS-based Adaptive Traffic Engineering (MATE) \cite{elwalid2001mate} & 2001 & Uses probing messages and explicit congestion notification packets to solve multipath routing which is formulated as optimization problem with the objective function of minimizing congestion on highly utilized links in the network. It is an online framework.

\\

TeXCP \cite{kandula2005texcp} & 2005 & Uses explicit congestion notification packets to perform flow splitting in order to balance load in real-time manner in response to actual traffic demands/ failures. It is an online distributed protocol.

\\

Common-case Optimization with Penalty Envelope (COPE) \cite{wang2006cope} & 2006 & Establishes optimal multiple paths efficiently for expected network scenarios while providing worst-case guarantees. It is a hybrid of oblivious routing and prediction-based optimal routing---a class of TE algorithms. \\

DATE \cite{he2007towards} & 2007 & Incorporates both congestion control and routing mechanisms to the enable \textit{edge routers} to split traffic for each source-destination pair over multiple paths. It is a cross-layer multipath routing framework.

\\

REPLEX \cite{fischer2006replex} & 2006 & Avoids oscillations which is inherent to load-sensitive routing. It is an online traffic engineering technique based on \textit{Wardrop} routing policies \cite{wardrop1952road}. Since it is not restricted to MPLS-based solutions, it is a more general solution compared to TeXCP and MATE. \\

Flowlet-aware Routing Engine (FLARE) \cite{kandula2007dynamic} & 2007 & Splits a single TCP flow across multiple paths without resulting in packet disordering. \\

Hedera \cite{al2010hedera} & 2010 & Provides a centralized load-aware scheduler for datacenters. \\

Path RE-Feedback with Loss EXposure (PREFLEX) \cite{araujo2010mutualistic} & 2010 & Uses estimates of loss rate to balance congestion by automatically splitting traffic. \\

MicroTE \cite{benson2011microte} & 2011 & Exploits short-term, partial predictability of traffic matrix in order to adapt to traffic variations. It is a load-aware centralized TE solution designed for \textit{datacenters}. \\

MP-TCP \cite{ford2011architectural} & 2011 & Shifts traffic from one path to another; hence, it is a load balancing scheme for end hosts. \\

Adaptive Load Balancing Algorithm (ALBAM) \cite{zhong2011adaptive} &  2011 & Performs multipath scheduling of mobile users who access heterogeneous wireless networks using multiple interfaces.\\

Hop-by-hop Adaptive Link-state Optimal (HALO) \cite{michael2013optimal} & 2013 & Solves the TE intradomain routing problem through adaptation to traffic demands and computing link weights locally and optimally using the same information as OSPF. It is the \textit{first} optimal link-state hop-by-hop routing algorithm.
\\
TRUMP \cite{he2008multiple} & 2013 & Proposes four adaptive and optimal distributed solutions using decomposition techniques to exploit mulitple paths.\\
  \\

\\
\multicolumn{3}{l}{\textit{\textbf{Theoretical Analysis}}}\\
\multicolumn{3}{l}{\textit{(Multipath algorithms for optimal flow splitting \underline{given} a set of paths)}}\\

\\
Kelly et al. \cite{kelly1998rate} & 1998 & Presents and analyzes two classes of rate control algorithms to set prices of utilized resources fairly: one algorithm from the perspective of user and another one from the network. \\
Wang et al. \cite{wang2001internet} & 2001 & Shows that shortest-path routing with positive link weights can be used to support arbitrary flow splitting to provide an arbitrary set of loop-free routes. This Internet TE problem is considered without using overlay networks. \\
Kelly et al. \cite{kelly2005stability} & 2005 & Presents and analyzes a quick load sharing/balancing approach across multiple paths between a source-destination node pair based on end-to-end measurements.\\
Han et al. \cite{han2006multi} & 2006 & Presents and analyzes the concept of \textit{multipath TCP} where congestion control and multipath routing are studied jointly. Using overlay routers, which support source routing, multipathing in networks is enabled. \\

\\
\multicolumn{3}{l}{\textit{\textbf{Theoretical Analysis}}}\\
\multicolumn{3}{l}{\textit{Multipath algorithms for optimal (or near-optimal) decisions on \underline{both} routing path and flow splitting}}.\\
\\

Banner et al. \cite{banner2007multipath} & 2007 & Provides \textit{optimal} and polynomial-time algorithmic solutions to congestion control using \textit{K-path routing (KPR)} and \textit{restricted multipath (RMP)} problems, both of which are NP-hard, using acceptable approximation schemes.\\

\\
\hline
\end{tabular}
\end{table*}

\vspace{2mm}
\section{How to Use Multiple Paths?}
\label{sec:usage_mp}

A network-layer multipath framework subsumes both the control plane operations of multipathing (that involves construction of multiple paths) as well as the data plane operations of multipathing (that involves the use of the constructed multipath paths for forwarding the flow). After having spent a lot of time focusing on how routes are computed, we now turn to the (arguably more) important question of how to use these constructed paths. We note here that route computation task is typically considered a network-layer function, while the usage of resources---and in particular, congestion control through which end-user's transmitting rate is controlled to ensure that the overall network performance is not deteriorated due to network congestion---is considered a transport-layer function. The routing task is managed at the network layer through the interaction of routers while congestion control is managed at the transport layer at end hosts. These systems unfortunately work largely independent of each other resulting in an unwelcome tussle that arguably is holding back multipathing from achieving the immense benefits it can offer \cite{araujo2010mutualistic}.

We note here that while the layered model is a good tool for conceptualizing networking functionality, many important aspects of the Internet---issues such as reliability, flow control, congestion control, QoS, and indeed resource pooling through multipath---cannot be pigenholed into a single layer. The rigid bifurcation of multipath-related functionality into layers results in an antagonistic tussle in which each layer attempts to balance traffic and perform resource pooling myopically according to its own needs to the detriment of the overall performance. Following the maxim \textit{``layering is a good servant, but a bad master''}, we recognize that the overall success of multipathing requires solutions beyond those that are implemented at the network layer. In this paper, therefore, despite our focus on network-layer multipathing, we also describe related issues (such as flow splitting, load balancing, and multipath congestion control methods) that are meta-network-layer and typically considered in the territory of the transport layer.


In this section, we see how the multiple paths constructed by the routing schemes described in Section \ref{sec:cons_sel_of_routes} can be used for forwarding the packets. We will see next how the processes of \textit{flow splitting} and \textit{traffic engineering} together contribute to the act of load balancing that is crucial to reaping the benefits of resource pooling with multipath.

Considering first a single flow, we see that the packets of a single flow can either use multiple paths by splitting the traffic to concurrently use the diverse paths, or alternatively, the multiple paths may be used in a backup configuration where the primary path is used until a node/ link failure along the primary path upon which the routing defaults to the backup paths. The purpose of \emph{flow splitting} in concurrent configuration is to achieve the benefits of resource pooling using which a collection of multiple paths appears as a single virtual pooled link. An important question that arises is how to decide the ratio of flow splitting since both of the computed multiple paths may have very different properties (in terms of delay, bandwidth, congestion, etc.). There are various requirements that we expect from a reliable flow-splitting multipathing system:   there should not be a need for per-flow state to ensure scalability; the system should be tunable, dynamic, and should support heterogeneity in bandwidths/ delay.
In addition, the reordering of packets should be minimized (for which it is important to recognize distinct flows, and have mechanisms to mark packets as belonging to a particular flow that uses a certain sub-path). Coming now to multiple flows, and how their packets share the multiple paths available over the network, it becomes important to consider both the utilization of the network resources and the fairness of resource allocation. This is the prerogative of \emph{traffic engineering} which is typically handled inside the network by ISPs. 

There are five important questions relating to how the multiple paths are used that influence the organization of this section. 

\textit{Firstly}, how many of the available multiple paths should be used for transmitting the traffic flow? While higher number of paths can improve the performance, the management cost also increases with increased number of paths. 

\textit{Secondly}, an important question is whether the multiple paths should be used concurrently or in a backup configuration? Traditionally most techniques have used multiple paths in backup mode for its simplicity although increasingly concurrent usage is being adopted to improve performance. 

\textit{Thirdly}, should the flow splitting/ load balancing be static or dynamic? Since the congestion on different paths vary dynamically, the load balancing and flow splitting mechanisms should also be ideally dynamic. However, in practice static flow splitting/ load balancing schemes are also used due to their simplicity and low overhead. 

\textit{Fourthly}, how can we use multiple paths to perform load balancing through the diffusion and control of congestion? In particular, traditional congestion control schemes need to be adapted for use in multipath scenarios. 

\textit{Finally}, an important aspect of load balancing is who gets to control the process? and where? Traditionally, there has been a tussle between end hosts and the network for network control with traffic engineering typically implemented in networks while congestion control implemented at the end hosts. 

Apart from discussing these five questions (in the order presented) in their respective subsections, we will also present a subsection that summarizes the trends presented in this section. 


\subsection{How Many Paths to Use?}
\label{subsec:how_many}

While multipathing can provide significant performance and reliability benefits, the computation and maintenance of multiple paths also entails costs. It has been shown that utilizing \textit{all} the available paths provides only limited performance gains over an approach that utilizes a limited subset of all paths \cite{wang2011cost}. In fact studies have shown that, to obtain performance gain offered by multiple paths, one or two alternative paths, in addition to the traditional single path, can suffice \cite{wang2011cost}, \cite{cidon1999analysis} with multiple reservation algorithms potentially providing a gain in terms of connection establishment time which is attractive for interactive applications \cite{cidon1999analysis}.

In previous research, it has been shown through multiple studies that the bulk of benefits of multipathing can be reaped with only two paths, and that increasing the number of routes provide diminishing returns \cite{wang2011cost}. This finding is formally established in a widely cited paper by Mitzenmacher on randomized load balancing \cite{mitzenmacher2001power}. In another work, Akella et al. have also posited similar insights in an empirical study \cite{akella2003measurement} focusing on multihoming.  More recently, Merindol et al. have proposed a multipath routing technique that also relies on two precomputed paths.

Key and Towsley \cite{key2007path}, \cite{key2011path} have similarly shown that  not many paths are needed to reap the benefits of multipath. The chosen set of paths need to however be dynamically monitored for their quality to ensure that the best paths are chosen and retained while any poorly performing paths are flushed out and replaced. While the dynamic load balancing is done at a rapid time scale, the exploration of routes can take place at a slower rate. To reiterate, the network-layer functionality of computing routes and finding connectivity is done relatively slowly while the dynamic load balancing over the chosen paths is done through the end systems via the transport layer rapidly. 

It must be pointed out the flow size is an important factor in the choice of how many paths to use, and in deciding how to prefer different access technologies in multi-homed systems. For instance, it has been shown in a study conducted by Deng et al. \cite{deng2014wifi} that for short flows, there is no significant benefit of using multipath, and selecting the right access network technology is more crucial for high performance. 

\subsection{Concurrent or Backup Paths}
\label{sec:con_backup}

In \textit{single-path routing}, a router only maintains a single route to each destination, whereas in \textit{multipath routing}, multiple routes are maintained. While most common configurations in multipath routing are to maintain a single primary path along with alternative paths, it is also possible to use multiple routes in a \emph{backup} or \emph{concurrent} fashion. 


Both of these approaches have their own benefits. Depending upon a particular network application or a specific metric (e.g.,  reliability or load balancing), a routing algorithm adopts any one of these approaches. To ensure reliability alongside load balancing in a network, the concurrent multipath approach \cite{leung2001mp}, \cite{zhang2008designing} gives the desired results. On the other hand, a set of multiple backup paths are precomputed in \textit{path failover} strategic algorithms \cite{chiesa2014exploring}, which are provisioned in order of a specific \textit{path ranking} mechanism, in the case of failure of the primary path \cite{marina2001demand}, \cite{nasipuri1999demand}. Information about the concurrency of multiple paths is summarized for various routing protocols in Tables \ref{tab:MultipathRouting} and \ref{tab:MultipathRoutingWireless}. 

Next, we discuss various aspects of backup and concurrent multipathing techniques in more detail in the following subsections.

\vspace{2mm}
\subsubsection{Backup Multipathing}
\label{subsec:backup mp}

Routing protocols that make use of multipathing with multiple backup paths usually construct a primary shortest path in a source-destination pair along with a set of alternative paths. The alternative paths in the set are maintained and updated according to the network topology changes. In case when the primary path fails, then the traffic is quickly shifted to one of the backup paths from the set of precomputed alternative paths. In this sense, the time incurred in route rediscovery is avoided, thus contributing to more reliable, fault-tolerant, and resilient routing.

Legacy protocols, particularly AODV, are improved to support multiple backup paths in AOMDV \cite{marina2001demand}. Similarly, DSR, which is another well-known routing protocol is enhanced in \cite{nasipuri1999demand} to improve the reliability of routing. In \cite{sambasivam2004dynamically}, multiple paths are provisioned in an adaptive manner by updating AODV to increase packet delivery ratio in networks. \textit{Static-fail-over routing} (SFR) is another important protocol making use of alternative backup paths to circumvent network failures.

Another work \cite{chiesa2014exploring} discusses the issue of reliability in multipath routing in the terms of \textit{backup paths}. In this work, SFR is proposed to study the positive and negative aspects of legacy \textit{fail-over routing} techniques.

\vspace{2mm}
\subsubsection{Concurrent Multipathing}
\label{subsec:concurrent mp}

With modern networking applications having huge data traffic demands, network congestion has become a significant problem. In legacy single-path routing protocols, many potential paths in a source-destination pair are ignored. In concurrent multipathing, alternative paths are exploited to relieve network congestion while improving network performance, efficiency, and reliability. 
A large number of works employing multiple paths, along with intelligent load balancing techniques, have been proposed in the literature. We present a representative sample of these works (many of which are based on concurrent multipathing) in Tables \ref{tab:MultipathRouting} and \ref{tab:MultipathRoutingWireless}.

We note here that utilizing multiple paths in parallel is not a recent idea. This general idea has been proposed under various names in the literature, viz. traffic dispersion, striping, and inverse multiplexing. \textit{Traffic dispersion} has a rich history in networking with Maxemchuk first proposing a traffic dispersing framework known as dispersity routing in 1975 \cite{maxemchuk1975}. This approach advocates spreading the traffic spatially (by dividing the original message into sub-messages and then transmitting them in parallel over disjoint paths of approximately equal length) to achieve load balancing and fault handling. The use of disjoint paths ensures that transmission errors on different paths are statistically independent. \textit{Striping} pools multiple resources to meet capacity requirements even if individual links/ network interfaces, or hardware components, cannot satisfy the required capacity \cite{adiseshu1996reliable}. \textit{Inverse multiplexing} refers to the phenomenon of aggregating channels to allow greater flexibility in selecting bandwidth \cite{fredette1994past}.

The concept of \textit{concurrent multipath transfer} is presented in the work \cite{liao2011introducing}. This concept implies that a source-destination pair that uses multiple paths in a multihomed environment can reap the benefits in terms of better network utilization and data transfer resiliency. This work addresses the issue of \textit{shared bottlenecks} in the computation of multiple concurrent paths. A metric based on \textit{path correlation} is proposed to resolve this issue. This metric tells us about the correlation of two paths in terms of shared bottlenecks. Lower value of this correlation metric indicates that the paths become more uncorrelated or disjoint.

Another important load balancing technique that is popular at the link layer in recent multipath-based topologies for DCs is \textit{Valiant Load Balancing} (VLB) \cite{valiant1982scheme, valiant1981universal, Valiant1990} (covered in Section \ref{subsubsec:vlb} in more detail). This technique has inspired many other multipath routing works to cater to the problem of congestion and inefficient utilization of network resources. An example of VLB-based work that uses uniform flow splitting is presented in \cite{zhang2008designing} in which a design of network backbone is provided to provision multiple concurrent paths. 

Recently, concurrent multipathing has become very popular and mainstream with the development of multipath transport protocols such as MP-TCP. The great promise of concurrent multipath transmission using transport protocols is the promise of true resource pooling using \textit{coupled congestion control} \cite{raiciu2011coupled} with which it will be possible to shift traffic away from congested paths. Recent fluid-flow models \cite{kelly2005stability}, \cite{han2006multi} have shown the benefit of shifting traffic from congested to uncongested paths. In particular, it was shown that multipath congestion control can perform the task of load-dependent routing stably at shorter time scales (at the same scale of TCP congestion control that operates at the order of RTTs). In effect, multipath congestion control assumes a role typically associated with network-layer routing (of moving traffic to another path thereby avoiding congestion hotspots). This motivates the discussion of multipath congestion control in our paper whose focus is on network-layer multipathing solutions. In a work related to concurrent transport-layer multipathing, Wischik et al. have proposed the principle of equipoise---or equally balanced concurrent multipath transport---as a concurrent multipathing technique that allows better resource pooling, resilience to faults on any subpath,  and better handling of traffic surges \cite{wischik2010balancing}.  

\subsection{Load Balancing Approaches for Multipath}

Broadly speaking, load balancing and flow splitting approaches can be either static/quasi-static or dynamic/load-aware. The static/quasi-static methods are suited for network traffic matrices that do not change rapidly, while the dynamic/ load-aware methods are suited for changing network traffic matrices. There are also a few theoretical works that describe optimal, or near-optimal, methods for load balancing and flow splitting. We have summarized major load balancing works, with their references, in Table \ref{tab:LoadBalancing}. 

\subsubsection{Static/ Quasi-Static Load Balancing}
\label{subsubsec: static_appraoch}

\begin{figure*}
\centering
 \begin{subfigure}[ht!]{0.3\textwidth}
                \includegraphics[width=\textwidth]{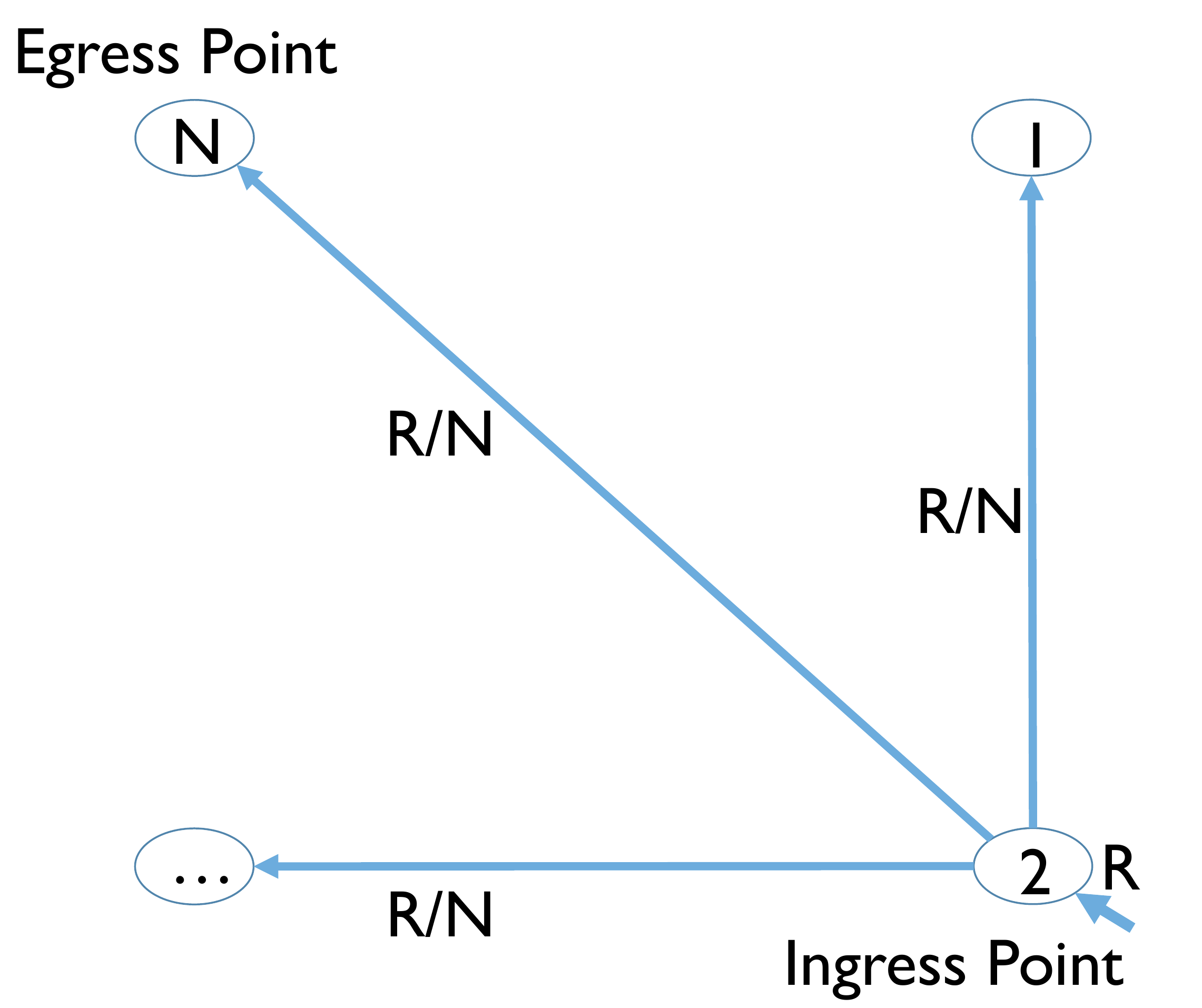}
                \caption{VLB Stage 1}
                \label{fig:vlbs1}
        \end{subfigure}%
        \begin{subfigure}[ht!]{0.33\textwidth}
                \includegraphics[width=\textwidth]{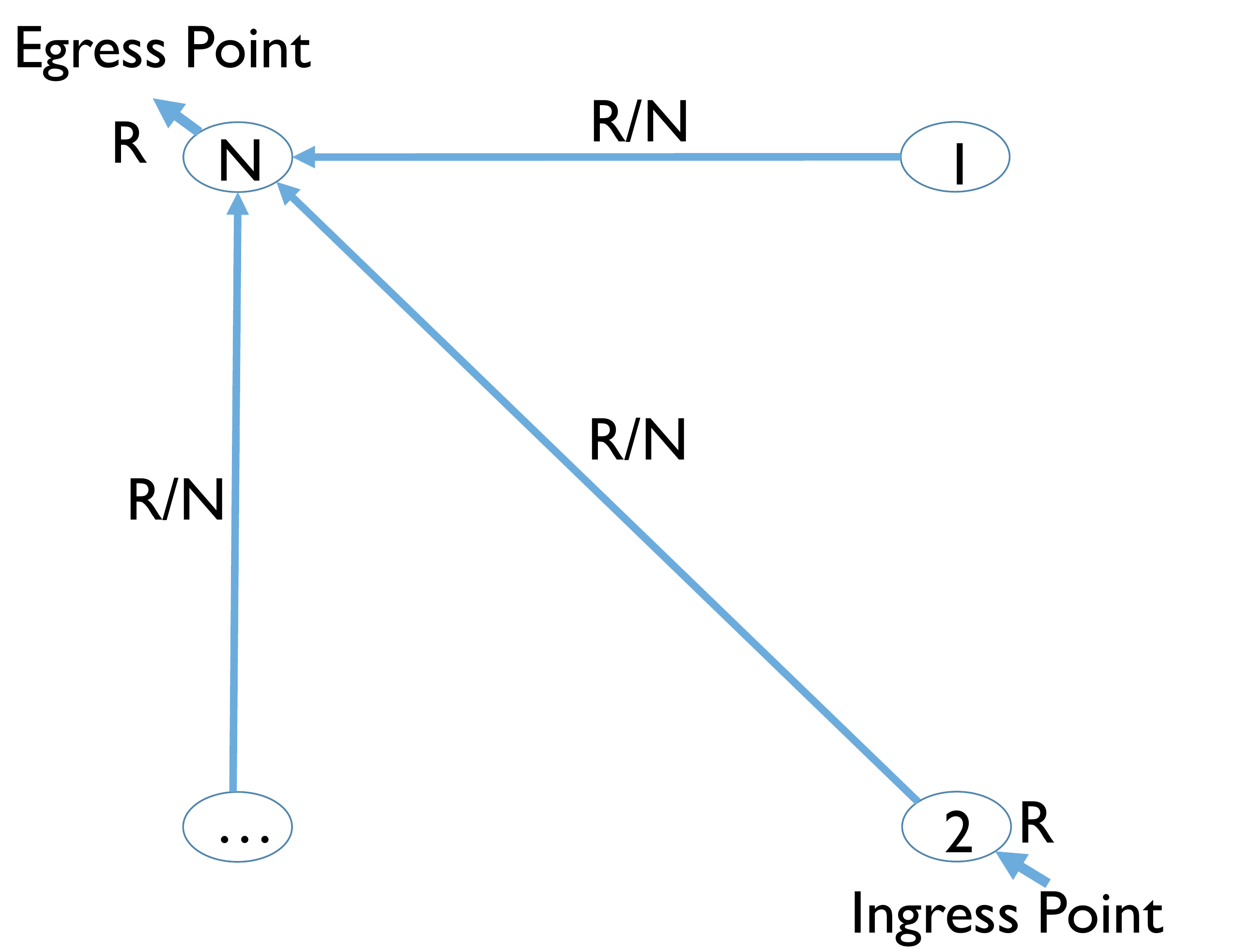}
                \caption{VLB Stage 2}
                \label{fig:vlbs2}
        \end{subfigure}
        \begin{subfigure}[ht!]{0.32\textwidth}
                \includegraphics[width=\textwidth]{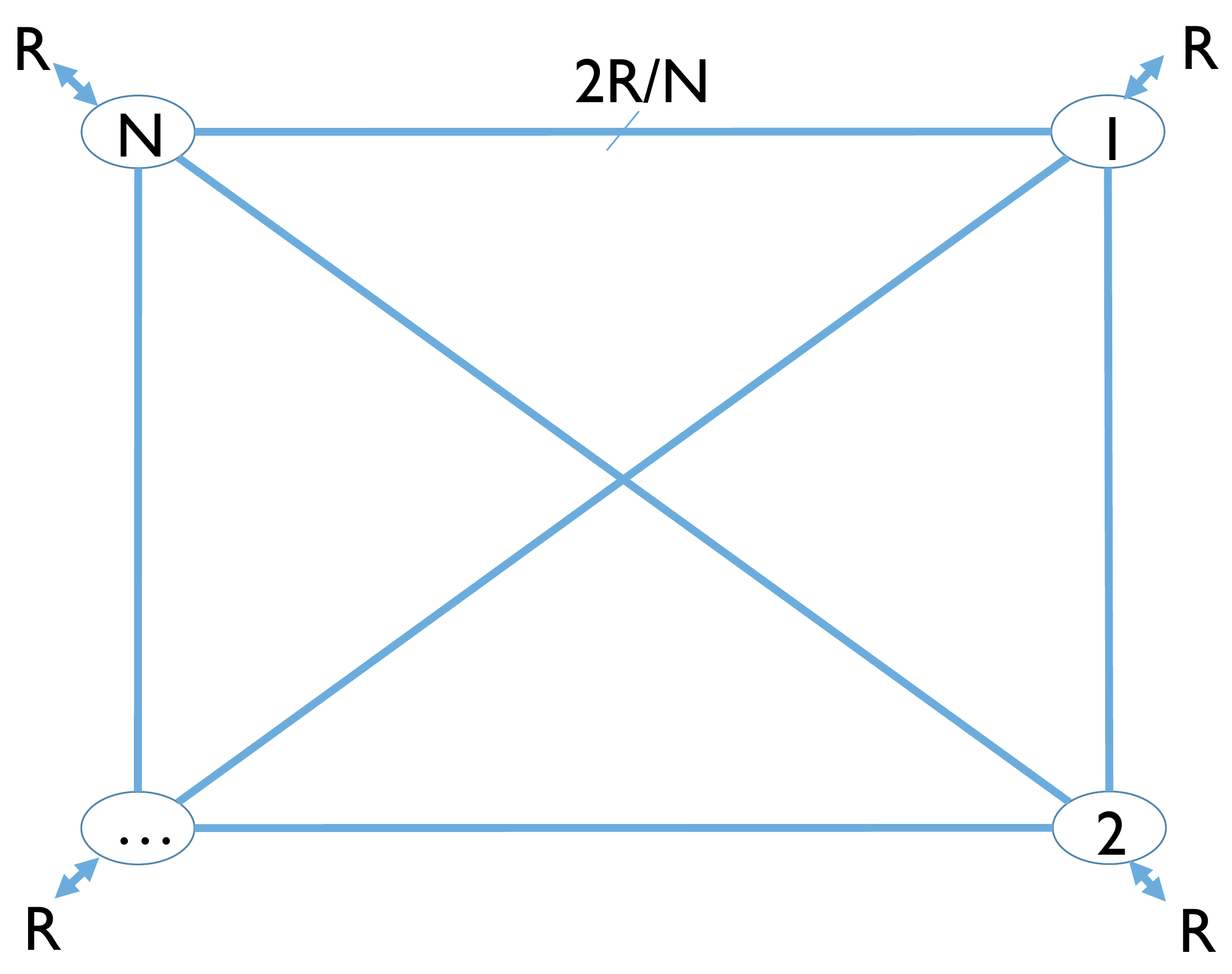}
                \caption{VLB Stage 1$+$2}
                \label{fig:vlbs12}
        \end{subfigure}
        \caption{Load balancing Through VLB}\label{fig:vlb}

\end{figure*}

This approach is adopted when traffic matrices are relatively stable with changes occurring after a significantly long period of time. Flow splitting techniques for load balancing in such networks are managed in a static or offline manner. Most of the static load balancing approaches aim to fairly schedule the flows giving equal preference to individual links \cite{shreedhar1995efficient}.

 \vspace{2mm}
\paragraph{Equal Cost Multipathing (ECMP)}
\label{subsubsec:ecmp}

ECMP \cite{moy1998ospf, hopps2000multipath, hopps2000analysis} is a multipath routing mechanism that has been employed in many routing protocols like OSPF and ISIS \cite{villamizar1999ospf, fortz2000internet}. Intuitively speaking, ECMP can be seen as a simple ``bonding'' technique that can load balance over equal-cost paths. A router can implement three main techniques while choosing a next-hop node to forward the traffic including the following.

\begin{itemize}

\vspace{1mm}
\item{\textit{Hashing:}} In this technique, a \textit{hash function} is applied to the header of each packet. The output of this hash function determines over which path this packet is forwarded to. In \cite{cao2000performance}, hashing is proposed for the purpose of network traffic load balancing over multiple equal-cost paths. This work makes use of a 16-bit CRC over the five tuples of headers of incoming packets. Five direct and one table-based hashing methods are presented. This work also presents a solution for keeping the packets of a TCP flow in order when it is split over multiple paths because of hashing.

\vspace{1mm}
\item{\textit{Per-Packet Round Robin:}} Fair queuing methods are deployed on a per-packet fashion to provide network resources to multiple contending queues of data packets in an efficient manner. In \cite{shreedhar1995efficient}, efficient and fast running algorithms for fair queuing, in which each flow that traverses across the network gets equal share of network resources, such as weighted round robin, are proposed. These can be utilized to load balance the incoming contending flows over the equal-cost multiple paths.

\vspace{1mm}
\item{\textit{Destination Prefix-based Approach:}} In this technique, different paths are attributed with different prefixes of a destination. The load is then split according to these prefixes \cite{villamizar1999ospf}.

\end{itemize}

Since ECMP deals with \textit{`equal-cost'} multipathing, it does not consider suboptimal routes although they can be potential candidates. \textit{Constrained Shortest Path First Protocol} (CSPF) \cite{manayya2010constrained}, which is an extended shortest path protocol, calculates path that fulfills a certain set of constraints, such as minimum bandwidth, maximum number of hops, and maximum end-to-end delay requirements.

\vspace{2mm}
\paragraph{Valiant Load balancing (VLB)}
\label{subsubsec:vlb}

VLB is an efficient parallel communication technique first presented by Valiant \cite{valiant1982scheme, valiant1981universal, Valiant1990} for general-purpose computer processors. VLB is an attractive form of load balacing since it can be applied on a per-flow or on a per-packet fashion, and can help build networks that can efficiently and reliably support any traffic matrix. In the case of VLB, it is sufficient to know the capacity of the ingress router of a network instead of the knowledge of the whole network traffic matrix. 

In VLB, each data flow is routed towards its ultimate destination in two stages: in the first stage, the data flow is routed to a randomly chosen intermediate router without the consideration of an underlined traffic matrix; the intermediate router routes the data flow to its ultimate destination in the second stage.

We explain the concept of the two-staged VLB operation with the help of a diagram shown in Figure \ref{fig:vlb}. Consider a network of $N$ nodes each with a capacity of $R$ as shown in Figure \ref{fig:vlb}. The two stages work in harmony as follows:

\begin{itemize}

\vspace{1mm}
\item{\textit{Stage 1:}} In the first stage, the incoming traffic that flows through the \textit{ingress point} is divided evenly, regardless of the eventual destination, among all the points in the network. As the capacity of the ingress point is $R$, and if it receives the traffic at this same rate, then the traffic is divided among the links giving $\frac{R}{N}$.

\vspace{1mm}
\item{\textit{Stage 2:}} In the second stage, each intermediary, which has received the traffic in the first stage, forwards the traffic towards the actual destination. As in the first stage, each node receives $\frac{R}{N}$ amount of traffic; so in this stage, each link is loaded with $\frac{R}{N}$ traffic by each node.

\end{itemize}

The combination of these two stages thus leads to the link capacity of $\frac{2R}{N}$ each for the normal/ efficient operation of VLB. Here, it can be noted that, if we only know the capacity of the ingress point (assuming the capacities of all the other points are equal), then there is enough information to perform load balancing, in a congestion-free way, on an incoming traffic load. However, an ingress router must not be overloaded. Uniform VLB---where incoming traffic is split equally at the ingress point over all the outgoing links---can promise 100\% network throughput. 

As an example application of VLB, Zhang et al. have used VLB to design a backbone network with fault tolerance \cite{zhang2008designing}. It has been discussed that, by 10\% over-provisioning of network elements in terms of nodes, a 50-node network continues to operate in a congestion-free manner with up to five node or link failures when VLB is applied to balance traffic load in such a network. Over-provisioning can be calculated by the ratio of an arbitrary number of failures to the total network nodes. 


\vspace{2mm}
\paragraph{Other Static Approaches}

MPLS with \textit{optimized path} (MPLS-OMP) \cite{villamizar1999mpls} is used for load balancing purpose for label-switched paths. TE, which is further explained in Section \ref{subsec:TE}, is used to balance traffic load across the label-switched paths. In the work of Fortz et al. \cite{fortz2000internet}, OSPF/ ISIS is deployed with known traffic matrix to perform TE in a near-optimal fashion under the constraints of shortest-path routing and even split of flows over equal-cost multiple links. In the follow-up works for changing network traffic matrix, Fortz et al. in \cite{fortz2002optimizing, fortz2002traffic} introduced a family of algorithms that adjust the link weights according to the changes in the traffic matrix. Here, OSPF/ ISIS is used for the traffic engineering purpose again. The fact that link weights are altered according to the changes in the traffic matrix makes the works quasi-static in \cite{fortz2002optimizing, fortz2002traffic}. In these works, the changes in the network traffic matrix are predicted for an intradomain routing. In another work \cite{sridharan2005achieving}, a TE solution for OSPF/ ISIS shortest-path routing is deployed for flow splitting purposes. The proposed solution computes approximately optimal link loads that provide a provable performance bound. In \cite{fong2005better, xu2007deft, xu2011link}, improvements for OSPF are presented so that non-shortest paths can be included for flow splitting. However, it is observed that a delay penalty that increases exponentially is incurred with the introduction of non-shortest paths.

\vspace{2mm}
\subsubsection{Dynamic (Load-aware) Load Balancing}
\label{subsubsec: dynamic_approach}

In this approach the TE solutions split the input traffic in an adaptive or online manner \cite{elwalid2001mate, kandula2005texcp, wang2006cope, he2008multiple}. The flow split is governed by the changes occurring in the network traffic matrix. We note that most of the works \cite{elwalid2001mate, kandula2005texcp, he2007towards}, which adopt this approach, address the problem of congestion control alongside balancing the input traffic load in a real-time manner.

Load-sensitive routing is prone to collisions. A general solution called REPLEX is proposed in \cite{fischer2006replex} to dynamically balance the traffic load and also avoid collisions. Another issue that can arise when the traffic load is split over multiple paths is the disordering of packets of a TCP flow. FLARE \cite{kandula2007dynamic} is proposed with real-time ability to split a TCP flow that does not result in disordering of TCP flow packets. MP-TCP \cite{ford2011architectural} is another online flow distribution method that renders an end user to shift the load from one path to another according to the input load demands and congestion feedback from the network. HALO is proposed in \cite{michael2013optimal} as the \textit{first optimal LS intradomain routing protocol}. It also deploys TE solution for load-aware traffic load balancing. Adaptive load balancing algorithm for multipath scheduling (ALBAM) is proposed for heterogeneous wireless access technologies for mobile users.

DCs are an important application area in which multipath routing with efficient load balancing mechanism provides efficient and data intensive operation. Centralized control solutions, which manage and schedule the DC traffic flows, are presented in the literature. Hedera, which is a central load-aware scheduler, is proposed in \cite{al2010hedera} for the DCs. This solution schedules flows in a dynamic and adaptive manner to efficiently utilize network resources. Another work for the same purpose, which is called MicroTE, is presented in \cite{ford2011architectural}. This technique is also load-aware and it uses short-term traffic matrix predictions to distribute the network traffic over multiple paths. Here, the short-term traffic, which has been shown to exist in DCs in this work, is used to infer unpredictable traffic in order to ameliorate congestion.

In \cite{araujo2010mutualistic}, a mutualistic resource pooling architecture named PREFLEX was proposed in which the hosts and edge routes worked in unison to collectively share the burden, and conversely reap the accrued rewards, of balancing traffic over multiple paths. PREFLEX is called a mutualistic framework since it allows congestion control (performed at the transport layer at end hosts) and traffic engineering (performed inside the network) to coexist and evolve independently by exposing the apparent network preferences and the transport expectations.

As we shall see, transport-layer solutions to load balancing are increasingly becoming more popular \cite{key2007path,honda2009multipath,wischik2010balancing,peng2015multipath}. Multipath transport protocols like MP-TCP already incorporate congestion control features that can automatically shift traffic from congested paths onto uncongested paths \cite{raiciu2011coupled}. As we discuss next, in many aspects transport-layer load balancing is very appealing. 

\subsubsection{Theoretical Results of Optimal Flow Splitting}
\label{subsubsec: theor}

\vspace{4mm}
\begin{quotation}
\textit{``Somewhere ages and ages hence:\\
Two roads diverged in a wood, and I---\\
I took the one less traveled by,\\
And that has made all the difference.''---Robert Frost}
\end{quotation}

\vspace{2mm}

It has been shown in multipath routing literature \cite{kelly2005stability} that, consistent with Frostian wisdom, taking the less-travelled-by path does indeed make all the difference in the context of multipath routing in communication networks. In multipath terminology, this means that multipath-capable flows should be designed to shift all their traffic from congested paths to uncongested paths. Interestingly, it has been shown that this load shifting can be done stably by the end hosts at the transport layer at a rapid time scale (at the order of RTTs) \cite{kelly2005stability}. In this section, we consider theoretical works that have analyzed congestion control and the benefits of flow splitting on multiple paths using fluid flow models. 

In optimal flow splitting, we are provided a set of paths, and we have to decide on what is the most fair and effective way to split traffic on the given set of paths. Fairness is an important goal to strive for and it entails providing every flow a ``fair'' share of the available bandwidth whenever there is congestion at a bottleneck. The most common form of fairness has been max-min fairness in which all the connections get the same share of the bottleneck. Another common form is proportional fairness in which any change in the distribution of rates will result in the sum of the proportional rates being negative. The rate control of TCP---which is based on additive increase and multiplicative decrease (AIMD)---is proportionally fair. In weighted proportional fairness, on the other hand, each connection is associated with a price and the amount paid per rate (and not the rates) is made fair. As an example, a connection with a price of two will get the same rate as two connections with a price of one each. 

It has been known through the work of Kelly et al. \cite{kelly1998rate} that, in a network in which each user chooses a price that maximizes its utility, the system evolves to a state that maximizes the total network utility. Two classes of control algorithms---one from the user's perspective and another one from the network's perspective---are presented to set prices for the utilized resources in a fair manner. 

In a follow-up work, Kelly et al. \cite{kelly2005stability} considered the important problem of joint routing and rate control where it was demonstrated that load sharing among multiple paths, in a multipath routing scenario, can be achieved in a similar time scale in which rate control is performed. But this quick load sharing approach must be performed while keeping in view the network stability (responsiveness vs. stability). This is why, in this work, a method is proposed to analyze end-to-end stability in the joint routing and rate control scenario with the presentation of a sufficient condition for this purpose. This sufficient condition works in a distributed manner and deals with the stability of end-to-end algorithm for joint routing and rate control.

Han et al. have also considered the problem of joint multipath routing and congestion control with multipath-TCP \cite{han2006multi}. This problem is studied based on an overlay network of routers that support source routing. Wang et al. in \cite{wang2001internet} have also considered arbitrary traffic split among multiple paths in a shortest-path routing scenario. But here, an overlay network is not considered in this TE approach. Links with positive weights have been used to support the distribution of arbitrary input traffic matrix.


\subsection{Multipath Congestion Control}


Apart from the issues of route computation and traffic engineering, another important issue is \textit{congestion control}. In practice, congestion on the Internet is passively controlled at the end hosts through TCP that observes and responds to packet losses. Alternatively, the network nodes can also actively manage their queues using active queue management (AQM) technology---using algorithms such as random early detection (RED) \cite{floyd1993random}, explicit congestion notification (ECN) \cite{floyd1994tcp}, or controlled delay (CoDel) \cite{nichols2012controlling}---to take preventive measures to avoid congestion.


The purpose of multipathing is to enable resource pooling. For effective resource pooling, it is important to develop the abstraction of a \textit{single-pooled resource} from the collection of diverse resources. The key to the creation of this abstraction is effective load balancing. More relevant to multipathing, load balancing can be performed over multiple resources (e.g., the load is balanced over multiple links/ paths). With multipath resource pooling, the aim is to pool multiple paths and to provide the illusion of a single much more powerful path. This resource pooling can provide the benefits of higher throughput, improved fault tolerance, and the capacity to handle traffic surges. 

There is a lot of ongoing work in the research community in developing efficient \emph{multipath congestion control}  \cite{raiciu2009practical}, \cite{key2011path}, \cite{wischik2011design}. This interest is further buoyed by the ready availability of path redundancy and diversity in modern networking environments such as datacenter networking. For maximum efficiency, there should be a strong emphasis on the construction of holistic multipathing frameworks that jointly optimize the various facets of network-layer multipathing.


\vspace{2mm}
\subsubsection{Coordinated or Uncoordinated Control}
With multipath congestion control, an important consideration is whether to use \textit{uncoordinated} or \textit{coordinated} control. Uncoordinated control can be implemented in the current Internet (running unmodified TCP) in a straightforward fashion. This makes uncoordinated control attractive for application-layer solutions, and explains the use of such schemes by P2P frameworks. With coordinated control, on the other hand, the coordinated controller can better balance load when the paths are fixed.  Kelly and Voice \cite{kelly2005stability} and Key et al. \cite{key2007path} have studied congestion management in multi-flow environments. They have shown that congestion response for each subflow should be coupled to ensure stable and efficient rate allocation between the subflows. This is important both when a path fails and also when a new path appears. This has been studied using tools of game theory by Key and Towsley \cite{key2011path} by the formulation of a path selection game which studies the choices a user should make about the set of throughput-optimal paths to use seeking to maximize their own net utilities. The ensuing Nash equilibria is shown to be welfare-maximizing social optima for the case of coordinator controller and for the restricted case of uncoordinated controller that do not exhibit RTT bias (unlike TCP which does). For uncoordinated controllers with RTT bias, the Nash equilibria may not coincide with the social welfare-maximizing optimal. Coordinated controllers have also been shown in \cite{key2011path} to improve upon greedy least-loaded resource selection as proposed by Mitzenmacher.

\vspace{2mm}
\subsubsection{Congestion Exposure}

Today's traffic management protocols are typically suboptimal since they disregard (i) protocol interactions between congestion control, routing, and traffic engineering; and (ii) the tussle that emerges from independent network \textit{or} end-host control of congestion. The task of harmonizing congestion control, routing, and traffic engineering is challenging due to the different timescales they act on (short, medium, and long timescales respectively). In recent times, there is renewed interest in integrating routing and congestion control in a cost-effective fashion. The EU-funded Trilogy project \cite{abt2009trilogy}---named as tribute to the crucial trilogy of ``routing, congestion control and cost effectiveness''---aimed at improving the quality of the internal functioning of the Internet through cost-effective resource pooling and integration of the routing and congestion control functionalities.   

An important component of the Trilogy architecture aiming to plug a gap in the Internet's architecture is Re-ECN (re-feedback of explicit congestion notification)---which is an economics-based congestion pricing resource management framework that holds users/ applications for the congestion they create. It has been shown in previous work that resource allocation through the traditional TCP method can be gamed/ exploited (e.g., a user can open multiple TCP flows to gain disproportionately high throughput at a bottleneck link) since the rate control is performed at the end hosts. The network operators have limited visibility into the congestion in the network and their focus has been traditionally on managing utilization. The traditional focus on ``TCP-friendliness'' \cite{floyd1999promoting} and the focus on equal flow rates have been challenged recently and their fairness and desirability have been questioned \cite{briscoe2007flow}. 

An important protocol in the Re-ECN framework is the Congestion Exposure (ConEx) protocol that can also lead to better integration of end-host applications and service provider, and harmonious interworking of various stakeholders on the Internet \cite{briscoe2007flow}. ConEx attempts to fairly ``divide the resource pie'' and mitigates the indiscriminate abuse of the Internet by an atypical form of traffic policing that only pushes back users who are limiting other users' freedom. This is done by charging users according to the congestion they cause, rather than by the volume of traffic generated. The underlying idea here is that we should not limit the traffic volume generated by users unless the resource share of other users is being impacted. Thus, if there are more people who wish to send data, there is a higher price of transmission, and vice versa. Without such a system, the system can be exploited causing indiscriminate use of the Internet leading to tussles between ISPs and user applications such as P2P \cite{briscoe2007flow}. 

In ConEx, the packets carry an accurate prediction of the congestion they expect to cause downstream, which is visible to the ISPs/ network operators who can now act on the congestion signal being exposed to them by the end hosts \cite{moncaster2009need}. This information can then be used for a variety of purposes including congestion policy, traffic engineering, and accountability. ConEx encourages well-behaved applications that aid load balancing (e.g., by delaying the transmission of traffic from congested times to uncongested times) and can help facilitate better network economic pricing models. This can allow ISPs to charge users for congestion caused while also providing users incentives to fairly consume resources since they have to now pay for the congestion they are causing. There is also the scope of novel tussle-aware congestion management schemes that utilize information made available to policing of network congestion \cite{briscoe2005policing}.

\vspace{2mm}
\subsubsection{New Developments in Congestion Control}

As an interesting example of a new congestion control algorithm, that involve greater interaction between the network and the end systems, consider Low Extra Delay Background Transport (LEDBAT)---a protocol used by Apple and BitTorrent for large background transfers that account for up to an approximate 15 to 20\% of the the global Internet traffic---which works by limiting congestion induced by the LEDBAT flow itself in the network. There has also been work in machine-learning-based congestion algorithms that are customized for user and network preferences. In the ReMy work, proposed by Winstein et al. \cite{winstein2013tcp}, the computer-designed endpoint algorithms meet the defined preferences and assumptions about the users, application traffic, and network much better than human-designed algorithms (even those algorithms that deploy code inside the network). In other works, ReMy searches for the best congestion-control algorithm and optimizes the expected objective over prior assumptions. This work promises that we can retain the end-to-end design philosophy and work with a simple ``dumb'' network, and have a computer-designed algorithm operates at end points and outperforms even the best in-network algorithm. 

\subsection{Which Layer/ Where to Perform Load Balancing/ Traffic Engineering?}

For the success of a load balancing/ traffic engineering scheme on the Internet, it is important that the proposed framework holistically integrates the various layers of the Internet architecture. The network alone is incapable of managing the traffic since the routers do not have enough information about the end-to-end traffic flow and thus cannot make an information decision about the paths to be used. The end host, on the other hand, does not know about the network traffic and cannot ensure fairness of the network alone. The problem with performing congestion control at the transport layer is that the control process may not have enough time to kick in for small flows. In many cases, application-layer multipathing is best suited to serve the unique needs of the application and can be used to choose from which server to download (as is done in P2P applications such as BitTorrent that already inherently utilize multipath). 

Our discussion on transport-layer multipathing in this paper, despite our network-layer focus, is motivated from the convergence of the research community on the important role that congestion-controlled multipath transport will play in the future multipath-enabled Internet. It has been argued in literature, following the work of Kelly and Voice \cite{kelly2005stability}, that end systems using transport-layer functionality, rather than the networking nodes implementing network-layer protocols, are best placed to implement load balancing---particularly across network domains---on the Internet. The information below the transport layer is too coarse to dynamically control the traffic and can only enable coarse-grained functionality (such as shifting of a few elephant flows away from hotspots within a network domain). Indeed, many application-layer protocols (especially P2P protocols such as BitTorrent) already perform multipath in their function. Implementing a generic multipath load balancing/ resource pooling functionality can help implement a new `narrow waist' of the future Internet. This also naturally fits with the existing role of TCP in implementing congestion and rate control functionalities. 

Broadly speaking, there are three main techniques for multipath resource pooling: (i) Routing-based traffic engineering; (ii) Application-based load balancing between multiple servers; and (iii) SDN-based approach to traffic engineering. 


\vspace{2mm}
\subsubsection{Router-based Traffic Engineering (TE)}
\label{subsec:TE}

Since it is common for large networks to offer significant path diversity, there is great interest in using TE to increase network efficiency. A common approach to implement flexible forwarding is to first classify packets (according to application requirements with packets for the same flow being similarly classified according to the `type of service' field in IP packet header) and then to map packets to different paths by having an edge router examines the packet header, evaluate the path characteristics, and determine a suitable path for a certain class of traffic. 

To forward packets using the alternative paths, routers that are part of the communication session establish a logical link with each other using tunneling. The tunneling approach can be utilized in MPLS or IP-in-IP tunnels. Although an extra header is needed for encapsulation; however, compared to the much longer prefix match of destination address, a label-based look up is an easier and faster approach. Using explicit routing, each router can specify a path using a packet header, or a label embedded inside the packet header. Tunneling is a useful approach for special applications that need end-to-end bandwidth guarantee. This approach is also employed in a way that a complete router-level path is specified along with IP options. Moreover, each intermediate node performs a label lookup to discover the outgoing link.


Traffic engineering using network-layer routing functionality is often performed at slow time scales and is not automated (often with a human in the control loop).  Internet traffic engineering functionality, like most of the Internet's control functionality, has emerged as a melting pot of numerous piecemeal fixes (such as the congestion control fix provided by Jacobson's work in TCP; the traffic engineering fix provided by MPLS; and the policy routing fix provided by BGP).

The BGP protocol, developed mainly for advertising policy-compliant loop-free connectivity, has limited support for traffic engineering. The cumbersome process of establishing peering, provisioning links, as well as tuning routing and QoS policies makes traditional router-based traffic engineering unattractive. This, along with the coarse-grained traffic engineering functionality by protocols such as BGP, has motivated the development of other approaches (such as the two approaches---application-based and SDN-based, traffic engineering---that we discuss next).

\vspace{2mm}
\subsubsection{Application-based Load Balancing/ Traffic Engineering}

The traffic engineering decision can be made by the application itself at the edge host. As an example, P2P applications like BitTorrent utilize multipath---but often in a fashion that is oblivious and impervious to the ISP (e.g., the peers are selected randomly without necessarily looking for peers that are local to the same ISP)---thereby impacting the efficiency and costs of ISPs. There has been some work done to allow the ISPs greater flexibility in managing the P2P applications. For instance, the P4P framework \cite{xie2008p4p} aims to equip the provider with the ability to manage P2P applications by performing smarter peer selection, better traffic distribution, and higher transfer speeds, all at a lower cost to the ISP. P4P does this by allowing the ISPs to notify the clients about the preferred local clients. Similarly, the Ono system aims to ``tame the torrent'' of cross-ISP data generated by BitTorrent by modifying the BitTorrent client to prefer local peers \cite{choffnes2008taming}.

While application-layer load balancing and traffic engineering can be quite effective, it can conflict with what the network operators are doing, causing an unfortunate tussle. In particular, application-layer multipathing, as performed in P2P applications, is suboptimal due to the lack of availability of topological information  to the application layer at the end hosts. The Application-Layer Traffic Optimization (ALTO) framework \cite{seedorf2009application} proposes to equip distributed Internet applications with sufficient topological information exposure to optimize resource utilization.  With ALTO, P2P performance can be improved by allowing better-than-random peer selection allowing P2P traffic to align itself with ISP constraints.

\vspace{2mm}
\subsubsection{SDN-based Traffic Engineering}

In recent times, SDN architecture has emerged as a viable platform to facilitate QoS-aware and efficient traffic engineering. A popular use case of SDN is to use traffic engineering aided by a centralized network-optimizing vantage point in datacenters, campus networks, and in service provider networks. 

The modern SDN trend---where network policies can be disseminated from a centralized controller to network elements such as switches---can help fundamentally transform TE as demonstrated most dramatically by Google with their traffic engineering system B4 \cite{jain2013b4}. B4 uses an OpenFlow-based SDN architecture that achieves nearly 100\% traffic utilization when splitting application flows across multiple paths while meeting application priority/ demands. SDN has recently used by McCormick et al. \cite{McCormick2014SDN} to provision real-time alpha fairness-based TE. The proposed algorithm executes in the range of milliseconds and can be used as an online tool.

\begin{figure}[!ht]
\centerline{\includegraphics[width=.5\textwidth]{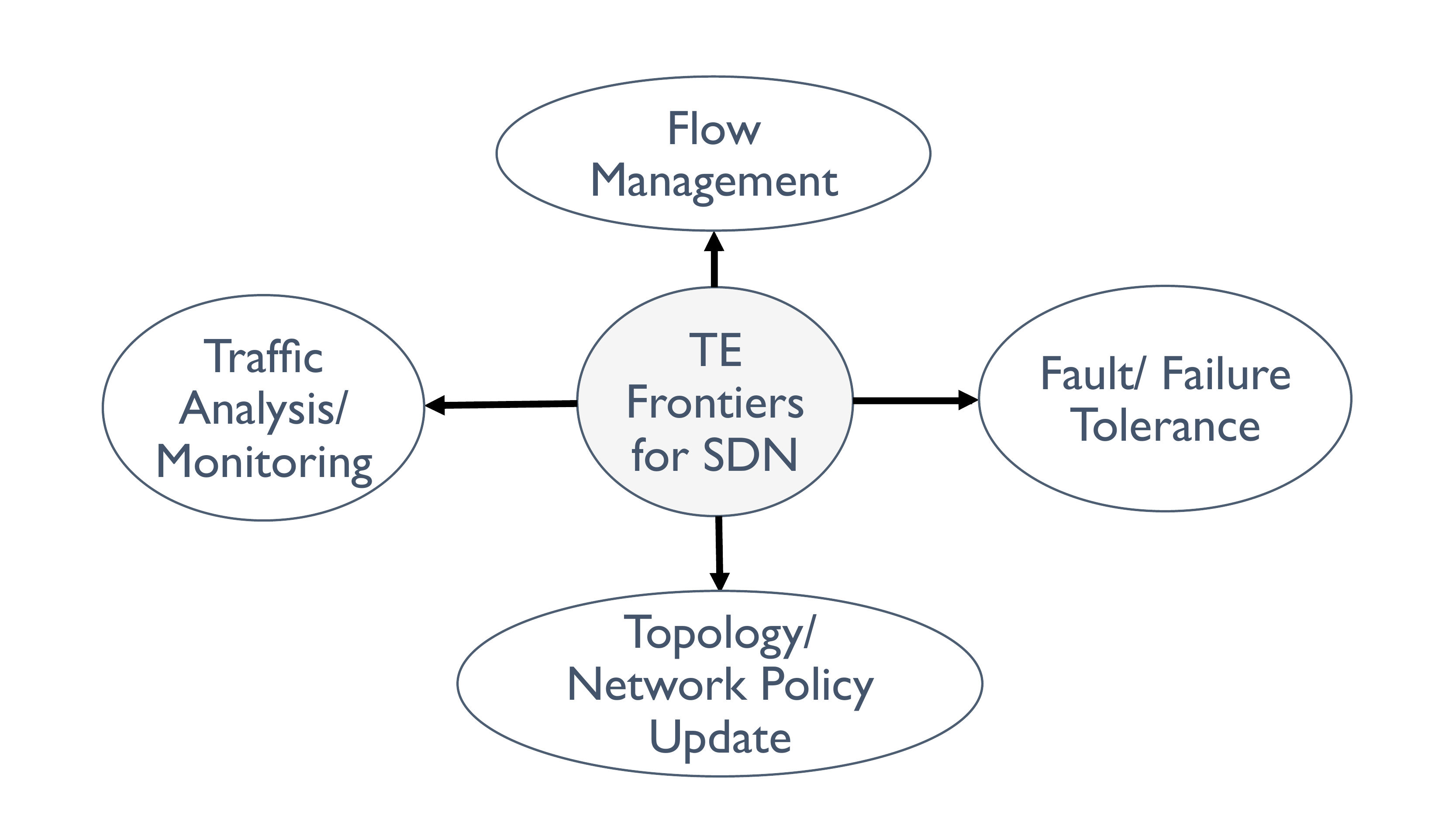}}
\caption{TE Frontiers for SDN (Source: \cite{akyildiz2014roadmap}).}
\label{fig:te_fron}
\end{figure}

An excellent survey for TE in SDN environment by Akyildiz et al. is presented in \cite{akyildiz2014roadmap}. We note that there are four frontiers of TE for the modern networking age of SDN. As mentioned in \cite{akyildiz2014roadmap}, these four frontiers are: (i) flow management, (ii) fault/ failure tolerance, (iii) topology/ network policy update, and (iv) traffic analysis/ monitoring. This is depicted in Figure \ref{fig:te_fron}. Here, we briefly describe these frontiers with the relevant reference works, which treat these frontiers in the perspective of SDNs and DCs.

\begin{itemize}

\vspace{1mm}
\item{\textit{Flow Management:}} In a SDN environment, each new type of flow is routed towards a SDN controller by an ingress switch. The controller then decides the path for this and the similar prospective flows. The controller disseminates this new path information to all the relevant switches. However, if there are a large number of new incoming flows, then it takes longer for a controller to decide new paths for each new category of flows, as well as to program the data planes of the relevant switches. This increases delay and can create problems, especially for the DCs, which have been solved using intelligent and innovative TE solutions. Hedera \cite{al2010hedera} is proposed to deal with these issues using a global view of active flows in the network. It creates non-conflicting paths for the flows as per the current routing and traffic demands in a real-time manner. Another similar work is Mahout \cite{curtis2011mahout}, which is an end host-based DC traffic management solution. In this study, traffic management is carried out by detecting large flows or \textit{elephant flows} at the buffers of the end hosts. Then, TE is applied to manage these flows using multiple paths with low control overhead.

\vspace{1mm}
\item{\textit{Fault/ Failure Tolerance:}} In a centralized control environment of SDN, failure recovery is a challenging task. In case of a failure, a controller must provide the affected switches or other network elements/ resources with a fail-over routing path(s). These paths can be considered as backup paths where the traffic is shifted once the primary path(s) fail. Provision of such type of alternative paths in a real-time fashion can be a challenging task for a SDN controller. Two types of failures can occur, so two types of failure recovery mechanisms are needed, namely (i) data-plane failure recovery and (ii) control-plane failure recovery. For data-plane failure recovery \cite{sharma2011enabling}, restoration and protection operations for network elements, like nodes and links, must be performed; while for control-plane failure recovery \cite{fonseca2012replication}, a single-point failure must be avoided. Due to the dependency on the controller in a SDN environment, backup and replication techniques for the controller failures are adopted.


\vspace{1mm}
\item{\textit{Topology/ Network Policy Update:}} Topology update in SDNs, unlike conventional networks, is related to the dissemination of network policy changes. The control plane of a SDN is responsible to plan for the updates in the network policy rules. The challenge is maintaining the \textit{consistency} of any updates in the policy rules in the data planes of all switches so that they are operating under the new rules. Problem arises during the \textit{update process} of the data planes of the underlying switches. Active flows, which are still under the influence of earlier network policy, can get affected or completely discarded during this update process. This phenomenon degrades the QoS of a SDN-based system. So, it is desired to perform the topology update mechanism in a near real-time fashion so that only minimum degradation occurs. Network policy update process is a planned mechanism by a SDN controller. Per-packet/ per-flow consistent and non-conflicting updates are ensured by solutions presented in \cite{reitblatt2011consistent, reitblatt2012abstractions}. In these works, the controller requires an ingress switch to label incoming packets or flows with a tag to tell the policy imposed on this packet so that the switches along a path can process accordingly. By this way, a single packet is processed according to only one policy---either the new or the earlier policy, but not both simultaneously.

\vspace{1mm}
\item{\textit{Traffic Analysis/ Monitoring:}} This frontier can be considered as a vital part for the whole TE mechanism in SDN. Accurate and efficient implementation of this frontier renders other frontiers to operate in a systematic cohort with high accuracy. In traffic analysis, different important observations---such as programming bugs, traffic data, network state and traffic/ data patterns---are taken. These measurements are important for timely detection of malfunctioning or the failures of network elements. Unfortunately, most of the current SDN-based systems still employ monitoring tools offered by the conventional IP domain systems. This approach often incurs huge delays, control overhead and inaccuracies in a SDN environment. A need for modern and SDN-related traffic monitoring tools is felt due to the aforementioned issues. Various traffic monitoring tools for SDN environment are available both from the industry, e.g, \cite{jain2013b4, hong2013achieving}, and the academia, e.g, \cite{wang2011openflow, sherwood2009flowvisor}. A few SDN-based traffic analysis/ monitoring works are discussed next as examples.  MicroTE \cite{benson2011microte} is a load-aware TE solution that predicts short-term traffic matrix that helps in adapting to the changing TE patterns. PayLess is another work that collects aggregate statistics at packet, flow and port levels in a real-time manner with high accuracy and low control overhead \cite{chowdhury2014payless}. 

\end{itemize}

\subsection{Summary of Trends}
\label{subsec:design_guidelines}

In this section, we summarize the major trends in how multiple paths are used in wired and wireless networks. We refer to Figures \ref{fig:infographicWired}
and \ref{fig:infographicWireless} for a visual depiction of the design choices made by various multipath protocols proposed for wired and wireless networks, respectively. We observe that multiple paths can be used in both concurrent/ backup configurations. It is observed (based on the works presented in this study) that wired/ general networks use backup and concurrent multiple paths almost equally, while in wireless scenarios, we see that the concurrent transmission option is adopted more often. Increasingly, the control of flow splitting, congestion control, and load balancing is being performed at the transport layer and the recent development of multipath transport protocols can be influential in strengthening this trend. With the recent proliferation of DCs, the use of dynamic/ load-aware flow splitting and SDN-based traffic engineering is also becoming more popular and will grow in importance in the future.

\section{Open and Current Research Issues}
\label{sec:future_wrk}


\subsection{Pragmatic Deployable Multipathing}

\vspace{2mm}
\subsubsection{More Flexible Support For Policies}


As the domain of applicability of multipathing expands with the growing popularity of multipath, the important problem of harmonizing the aim of efficient multipathing with the policies of the various stakeholders involved becomes more complex. The BGP protocol has traditionally provided limited support for traffic engineering. In particular, BGP does not natively facilitate common tasks such as balancing load across multiple links to a peering AS \cite{feamster2003guidelines}. Due to the wide deployment of BGP, a clean slate approach to interdomain multipath routing that breaks away significantly from the normal operation mode of BGP is unlikely to be deployed due to the central role of BGP on the current Internet. There have been various works proposed to introduce flexible interdomain multipath routing but the problems are far from fully resolved.

In addition, mulitpathing, by its very nature, brings to bear the possibility of policy complexity and conflict (both locally as well as globally). For example, a local conflict can arise between the objectives of energy efficiency and minimizing price when using WiFi/ Bluetooth/ Zigbee vs. Cellular \cite{vallina2013energy}; whereas example of a global conflict could be the need to respect the interdomain peering arrangements while performing multipathing efficiently \cite{xu2006miro}. On a mobile device, multipathing brings about interesting tradeoffs: e.g., if the desire is to plainly maximize battery life, then the lowest-energy interface (Wi-Fi or 3G) will be used; else if the interest is in maximizing throughput, then it will make more sense to use MP-TCP-like protocol to concurrently utilize both the interfaces; finally, if the preference is for price minimization, then a host of factors including the applicable pricing (whether roaming or not) will have to be incorporated. While there are a number of ways in which the end points could interact with the routing systems at the intradomain and interdomain levels to offset the policy tussle and ensure efficient multipathing, nothing concrete has been agreed upon yet, and this remains an open research issue.

\vspace{2mm}
\subsubsection{Seamlessly Handling Heterogeneous Paths}

By its nature, multipathing, especially in multi-homed devices/ networks, involves traversing heterogeneous technologies/ domains. As an example, Wi-Fi and cellular interfaces are pervasively available on modern wireless devices. Wi-Fi and cellular have very different characteristics: Wi-Fi typically exhibits stable RTTs with relatively high loss rates, while cellular has variable and large RTTs with low loss rates (with the large variable RTTs attributed to ``bufferbloat'' \cite{gettys2011bufferbloat}, \cite{jiang2012understanding} that refers to unnecessary provisioning of buffers throughout the network which hinders the congestion control process). The heterogeneous characteristics of the underlying technologies on the Internet can cause performance issues. It has been shown that MP-TCP performance can suffer when Wi-Fi and cellular flows are coupled due to cellular bufferbloat which causes the problem of \textit{flow starvation} and \textit{idle restart} \cite{chenbufferbloat}. More research needs to be done on how to tame the inherent heterogeneity on the Internet so that multipathing can show its full benefits. 

\vspace{2mm}
\subsubsection{Multipathing and Cluster-based Networks}

Despite the many advantages offered by multipathing, the high cost of maintaining multiple paths impedes wide-scale adoption. A potential solution to this problem is clustering in which nodes are grouped into logical groups/ clusters, with a clusterhead elected to represent the group. Clustering has many routing applications such as its ability to limit the amount of control messages exchanged, and its ability to aid abstraction (e.g., the Nimrod routing architecture \cite{castineyra1992nimrod} uses clusters to abstract the internal topology, hiding away the details and only revealing additional details upon request).  While clustering has been widely investigated, there is a lack of research work on supporting multipathing in clustered networks---particularly the effects of the underlying cluster structure (e.g., cluster size) and clustering operations (e.g., cluster maintenance) on multipathing. Further investigation can be pursued to investigate clustering as an effective solution to improve the scalability of multipathing.


\vspace{2mm}
\subsubsection{Interworking with Middleboxes}

Middleboxes are network entities that cause interruptions in end-to-end protocols like TCP. Middleboxes can be of various types and purposes, such as firewalls, network proxies, network address translators (NATs), and so on. These entities, unlike routers, are link layer entities that check the layer three information of a data packet. The proliferation of these middleboxes is one of the reasons for the ossification of the current Internet. With modern, often software-defined, networks and end hosts with potentially multiple interfaces/ addresses, the need for an end-to-end multipathing protocol is direly felt. The TCP has an inherent capability to support multipathing through the correct settings of the option fields in TCP packets. The major impediment in the way of realizing this multipath dream is middleboxes as they are tamed to work in an ossified fashion of processing packets assuming no prospective evolution. While making TCP to support multipath, it has to be considered that how the space of sequence numbers is to be used \cite{honda2011still, raiciu2012hard}. Do we need a single space for all the flows or a separate space will work just fine? If there are gaps in the sequence space of multiple TCP flows, then what might happen at the middleboxes? How will transmission, ordering and retransmission work in the presence of traditional middleboxes \cite{honda2011still, raiciu2012hard}? Future research can be pursued to investigate these open questions.

While traditional middleboxes can create problems for multipath TCP,  middleboxes can also come to the aid of MP-TCP. For example, Detal et al. have proposed a custom middlebox known as MiMBox that works as a protocol converter between MP-TCP and TCP \cite{detal2013multipath}. Since MP-TCP requires modification of the end hosts, it suffers from the chicken-and-egg deployment problem; MiMBox can facilitate the transition of the Internet's transport towards MP-TCP.

\subsection{Addressing Multipathing Holistically}

The Internet's control architecture is not a result of carefully planned organization of functionality. Such an ``architectural soup of network controls'' is ill equipped to support multipathing and there is a dire need of a holistic control framework for multipath routing. Some particular problems are described below.

\vspace{2mm}
\subsubsection{Cross-layer Multipath Support}

Besides the network layer, there has been work on exploiting the benefits of multipathing at other layers; hence it is important to deploy and exploit cross-layer optimization in order to develop effective Internet-wide multipath solutions. The network layer multipath solutions must interwork with application, transport and link layers as described below.

\vspace{1mm}
The interaction of the \textit{application layer} and the \textit{network layer} can help to create efficient multipath solutions. The development of overlay routing often requires cross-layer support. It is often the case that certain routing-related observations (such as communication failures at the application layer in which the network layer fails to detect) can only be suitably made at the end hosts. The end hosts can utilize the overlay networking technique at the application layer to implement multipath routing on top of today's relatively inflexible routing system. It is also often the case that datacenters use load balancers to direct flows at the application layer to suitable servers keeping in view the network conditions. These observations motivate greater support for cross-layer interactions between techniques at the network and application layers.

\vspace{1mm}
Previous research has also demonstrated the \textit{importance of the transport layer} in efficient multipathing. Kelly and Voice \cite{kelly2005stability} have shown that, while the network layer is well suited to provide structural information, the load balancing task is best managed at the transport layer. In particular, the key constraint on the responsiveness of each router is the RTT of that route (the information of which is conveniently available at the transport layer).  There is also a great need for increased \textit{interaction between the transport layer and other layers}. For example, transport layer can work with network-layer multipathing to reorder out-of-order packets arriving at the destination. There is thus a strong motivation of greater interaction between the network and transport layers for efficient multipathing. With recent research progress in transport layer protocols (with works such as MP-TCP, SCTP, along with numerous cross-layer solutions \cite{he2007towards} being proposed), the intersection of the network and transport layers for multipathing is expected to become a promising area ripe for further research. 

\vspace{1mm}
Finally, exploiting the information at the \textit{link layer} is also very important, especially for wireless networks where channel allocation and assignment are performed to encourage diversity \cite{mir2012unified}, \cite{farooq2013game}. It is noted here that in wireless networks, transmissions on orthogonal channels can take place simultaneously, and thus orthogonal channels can be considered as logically distinct channels/ paths. Multipath protocols can increase their efficiency by exploiting the multiplicity of routes made available through the use of channel-diversity in multi-channel wireless networks \cite{zeeshan2010backup}. In a similar fashion, the interface diversity (or the availability of multiple radio network interface cards) of multi-radio  wireless networks \cite{hassan2013quantifying} can be exploited to improve the performance of multipath protocols. 

\vspace{2mm}
\subsubsection{Mitigating the Tussle Between the Various Stakeholders}

Due to the fact that resource pooling mechanisms based on multipath are built on multiple layers and levels, and that a coherent framework addressing all aspects of multipathing (route computation, load balancing, congestion control, etc.) is lacking, it is important to ensure that the various resource pooling mechanisms work well in harmony and do not conflict. Towards this end, there needs to be greater synergy between the end systems, the end networks, and intermediate autonomous systems so that potential mechanisms/ policy conflicts (that can lead to suboptimal performance) can be avoided. 

As an example, P2P applications such as BitTorrent perform load balancing at the application layer independent of the ISPs by preferentially retrieving data on uncongested paths. Such an arrangement could come close to optimizing the cost for congestion pricing. However, with AS peering managed by the ISPs with arbitrary pricing models, the conflicting policies of the end systems and the ISPs can lead to a significant performance loss. It has been shown that the `cost of anarchy' resulting from the end systems and the ISPs adopting different metrics for congestion can potentially be arbitrarily high \cite{roughgarden2002bad},  \cite{acemoglu2007partially}. 

While network-layer multipathing solutions will play an important part in the future multipath-capable Internet, efficient resource pooling of multiple paths will eventually require cross-layer support and a seamlessly integrated resource sharing framework. It has been argued convincingly that the traditional inability of the Internet to progress beyond a single-path paradigm was less rooted in the lack of routing solutions but more so in the inflexible resource sharing model of the Internet which had an inevitable tussle between the hosts and the network \cite{clark2002tussle}. This tussle restricted the overall promise of resource pooling since the independent policies of the stakeholders often differ---usually at the expense of each other. Effectively, the various stakeholders are performing load balancing in a siloed fashion (independently and while trying to be inconspicuous to others). This situation is far from ideal since the interaction between the stakeholders will be at best a form of commensalism (a situation in which one entity benefits while the other parties remain unaffected) and at worst destructive for each other.

\begin{figure}

\centering
\includegraphics[width=.45\textwidth]{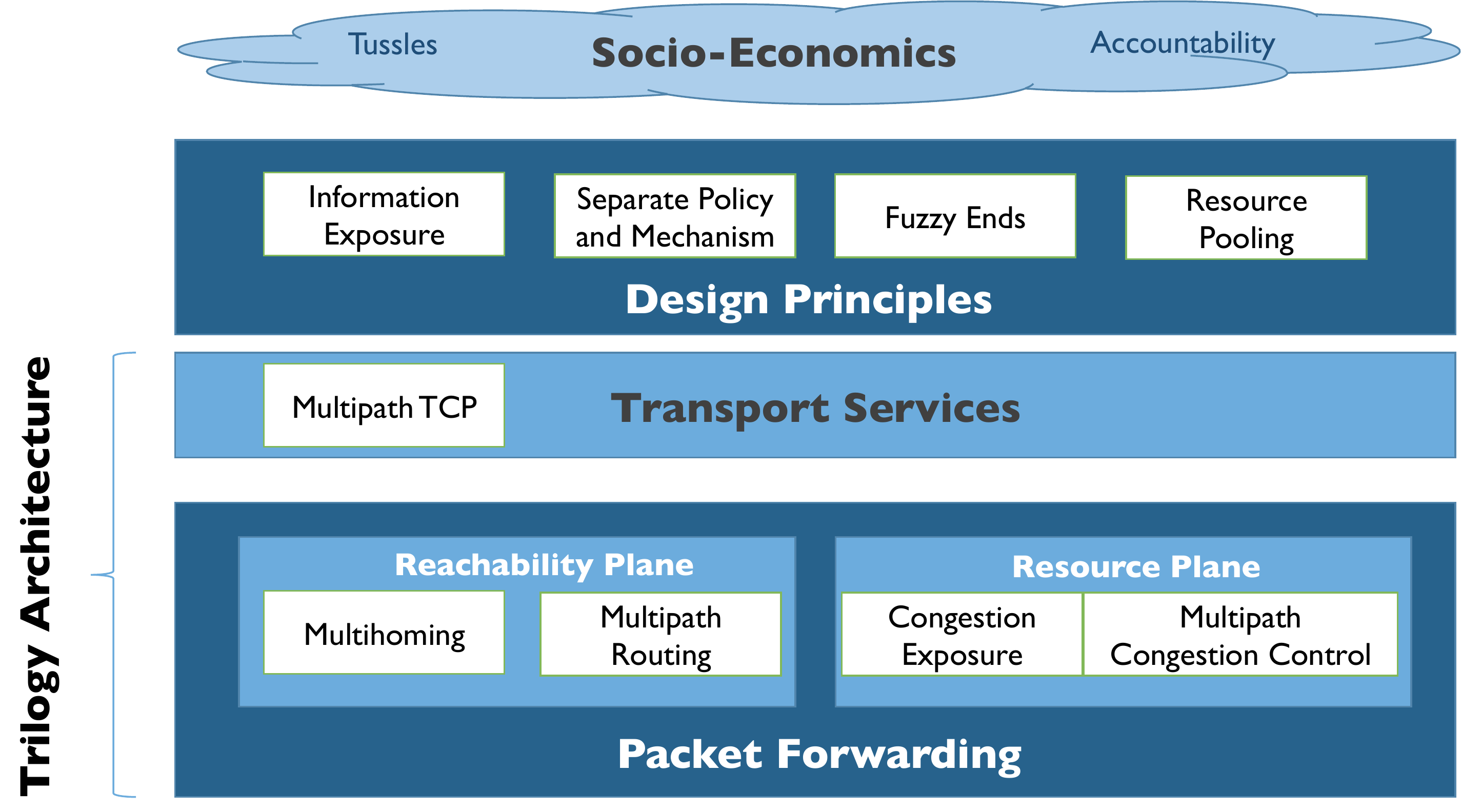}
\caption{A depiction of the holistic multipathing architecture proposed 
by the Trilogy future Internet architecture.}
\label{fig:trilogy}

\end{figure} 

In the EU-funded Trilogy project \cite{abt2009trilogy}, there has been an explicit attempt to acknowledge and manage the tussles between the different stakeholders in the network. Trilogy projects propose a new tussle-aware  Future Internet architecture that intrinsically incorporates multipath routing and transport while integrating technical solutions with socio-economic factors. Trilogy project proposed four general design principles for the future Internet \cite{ICT216372}, \cite{ford2009new}. Firstly, the \textit{Information Exposure Principle} requires the exposure of sufficient information about resource usage to support efficient and timely resource allocation. Secondly, the \textit{Separation of Policy and Mechanism Principle} proposes to separate higher-level policy decisions from standardized implementation mechanisms. Thirdly, the \textit{Fuzzy Ends Principle} proposes a mutualistic framework in which the  endpoints can delegate some functions to the network. Finally, the \textit{Resource Pooling Principle}---most relevant to multipathing---dictates that the network resources should be pooled effectively so that spare capacity can be used effectively to counter overload and faults.

The Trilogy architecture (shown in Figure \ref{fig:trilogy}) serves as a candidate embodiment of these principles. The overall architecture consists of a packet forwarding layer and a transport services layer. The packet forwarding layer comprises two players: (i) the \textit{reachability plane} which is responsible for hop-by-hop routing necessary for enabling network-wide reachability, and (ii) the \textit{resource plane} which is responsible for rate control and for deciding how to perform resource allocation between different packets/ flows. The transport services layer, on the other hand, manages the functionality of reliability, flow control, and message framing transparently from the packet forwarding layer. The transport layer is implemented end-to-end and interacts with the application layer. 



\textit{Multipath TCP} and \textit{Multipath Routing}, along with \textit{Congestion Exposure}, comprise the three building blocks of the Trilogy architecture.  Multipath TCP can directly exploit multiple paths on multi-homed endpoints whereas multipath routing allows the computation of disjoint paths even for single-homed devices (multipath theory indicates that the benefits of multipath can be reaped even without complete path disjointness as long as there is no shared path bottleneck). A synergy of transport and network layer multipathing is thus proposed in a holistic multipath solution that can allow faster traffic engineering (now performed end-to-end at the timescale of RTTs). Facilitating this interaction is `congestion exposure' which enables all nodes to see the congestion they cause by sending traffic. Congestion exposure also allows the operator to judge which paths are good and should be made visible to the end users, while also being able to see if the traffic engineering performed by the end hosts is satisfactory and not malicious. 

\subsection{Multipathing and Future Internet Architectures}

As explained in Section \ref{sec:intro}, we anticipate that multipathing will be a big part of the future Internet. Many of the future Internet architectures utilize multipathing in their framework. In this section, we will describe the role multipathing plays in the information-centric networking (ICN) and SDN architectures.

\vspace{2mm}
\subsubsection{Multipathing in Information-Centric Networks}

The ICN architecture employs content caching as an intrinsic and ubiquitous part of the networking infrastructure. A key aspect of ICN is to decouple location and content thus allowing content to be retrieved chunk-by-chunk from any location that is convenient. ICN inherently supports multipathing since a node's `Interest' for a content is not sent to a single location but can be sent through several interfaces with the replies being recomposed for application use. In addition to resource pooling of end-to-end resources and paths (as is done by multipath TCP protocols), resource pooling can be done on the basis of in-network resources (as is proposed in the ICN context with the in-network resource pooling work proposed recently \cite{psaras2014revisiting}) to improve reliability, end-to-end throughput, network efficiency, and flexibility.


\vspace{2mm}
\subsubsection{Using SDN for Multipath Provisioning}


More research needs to be conducted to discover how to best use the centralized paradigm offered by SDN, along with its more flexible control abstractions, for managing and provisioning multipath routing. SDN has been recently used by McCormick et al. \cite{McCormick2014SDN} to provide real-time alpha-fairness based traffic engineering. The proposed algorithm executes in the range of milliseconds and can be used as an online tool. Using the SDN architecture to split the control and data planes can also enable rapid response to network failures \cite{caesar2010dynamic}. In particular, in a SDN-based setting, multipath routing can not only improve performance and reliability, but can also forestall the need to dynamically recompute routes in response to failures.

\subsection{Deploying Diversity Efficiently}

One of the main benefits of multipathing is the pluralism and diversity afforded by the presence of independent links. It has been observed in a wide variety of settings that systems exploiting diversity and pluralism can outperform individual systems that do not rely on diversity \cite{surowiecki2005wisdom}, \cite{page2010diversity}. In this regard, the following are some important points to consider.

\begin{itemize}

\vspace{1mm}
\item \textit{Quantifying the Available Diversity}. Various quantitative measures of diversity have been developed in various sciences including entropy measures, variation measures, distance measures, and attribute-based measures\cite{page2010diversity}. Although it is unlikely that a single universal ``one-size-fits-all'' quantitative definition can be defined, there is a strong need of quantative measures of multipath diversity to build a rigorous theory on multipath routing. Some initial works on using diversity index in networking have been proposed \cite{dolev2010routing, chellappan2013application}, but there is a lot of scope for further work.

\vspace{1mm}
\item \textit{How to Use the Degrees-of-Freedom Afforded by Multipath?} The various degrees-of-freedom afforded by multipath can be used differently by application designers. For example, if we consider wireless communication at the physical layer, the availability of multiple paths can be used for either \textit{multiplexing} or for \textit{reliability} \cite{lozano2010transmit}. Similar options can also be used at the network layer, and the network designers should carefully evaluate how to use the available diversities.

\vspace{1mm}
\item \textit{Decorrelated and Disjointed Links/ Paths.}  To truly reap the benefits of multipath, the pluralism and diversity available through multiple paths should be leveraged. Care should be adopted to ensure the independence of the selected links/ paths.

\vspace{1mm}
\item \textit{Avoiding too much diversity.} The idea that diversity works well in moderation but can backfire when overdone is well known in common folklore. We know that ``two heads are better than one'' and also that ``too many cooks can spoil the broth''. While increased path diversity typically results in better performance and more robustness, there is generally a diminishing marginal utility with increasing diversity, and too much diversity can even potentially cause performance degradation \cite{page2010diversity}. In particular, coordination and management overheads (such as the control overhead of managing a large number of routing paths) can become prohibitively large with greater diversity. The implications of this need to be studied in the context of network-layer multipath solutions. 

\end{itemize}

\subsection{Multipathing and Green Networking}

The problem of energy efficient multipathing is an important concern for battery operated mobile devices. While mobile devices can increase their throughput by striping their connections over heterogeneous networks, this comes at a cost of higher energy consumption. With the proliferation of mobile devices, the incorporation of energy efficiency mechanisms into the design of multipath solutions has become extremely important. Early work on energy-efficient multipathing has already shown the potential for improving smartphone's energy consumption \cite{pluntke2011saving,raiciu2011opportunistic,paasch2012exploring,lim2014green}. 

Apart from its importance for wireless networks, energy efficiency is also critically important for datacenter networking (which operates at the scale of huge warehouses, where even minor savings can add up significantly). More research needs to be conducted in the area of energy efficient ``green multipathing'' to use the power of multipathing while reducing the energy consumption on the Internet. 

\section{Conclusions}
\label{sec:con}

In this paper, we have surveyed the existing literature on network-layer multipathing. After describing the motivation and benefits of multipath in networks, and establishing that the future of the Internet is multipath based, we discussed two important problems that are related to multipathing, namely the issue of how to compute the multiple routes to be used, and how should these multiple routes be used by traffic flows. While our focus is on network-layer multipathing, our coverage is holistic as we also discuss issues such as congestion control, flow splitting, and resource pooling (which do not fall neatly into a single layer). In addition to highlighting the main problems in network-layer multipathing, and describing the various approaches proposed, we have provided a broad-ranging survey of multipath protocols in different kinds of networks. Despite a vast amount of work on network-layer multipathing, many open questions remain; we conclude this paper by highlighting some important issues that require further investigation. 

\section*{Acknowledgment}

Arjuna Sathiaseelan and Jon Crowcroft were funded under the EU 7th Framework Programme, grant agreement number 317756, Trilogy II.

\bibliographystyle{IEEEtran}
\bibliography{multipathing}

\end{document}